\documentclass[usenatbib]{mnras}

\usepackage{newtxtext,newtxmath}

\usepackage[T1]{fontenc}
\usepackage{ae,aecompl}

\usepackage{graphicx}	
\usepackage{amsmath}	
\usepackage{amstext}
\usepackage[usenames]{xcolor}
\usepackage{booktabs,gensymb,comment}
\usepackage[referable]{threeparttablex}

\newcommand{\revv}[1]{#1}
\newcommand{\rev}[1]{#1}
\newcommand{\Msun}{\,M_{\odot}}

\newcommand{\epsff}{\epsilon_{\mathrm{ff}}}

\title[Stellar Clumps in High-Redshift Galaxies]{Origin of Giant Stellar Clumps in High-Redshift Galaxies}

\author[Meng \& Gnedin]{
Xi Meng\thanks{E-mail: xim@umich.edu}\href{https://orcid.org/0000-0002-8276-4164}{\includegraphics[scale=0.6]{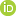}} and
Oleg Y. Gnedin\href{https://orcid.org/0000-0001-9852-9954}{\includegraphics[scale=0.6]{orcid.png}}
\\
Department of Astronomy, University of Michigan, Ann Arbor, MI 48109, USA
}

\date{Accepted XXX. Received YYY; in original form ZZZ}

\pubyear{2019}

\begin{document}
\label{firstpage}
\pagerange{\pageref{firstpage}--\pageref{lastpage}}
\maketitle

\begin{abstract}
We examine the nature of kpc-scale clumps seen in high-redshift galaxies using a suite of cosmological simulations of galaxy formation. We identify rest-frame UV clumps in mock HST images smoothed to 500~pc resolution, and compare them with the intrinsic 3D clumps of young stars identified in the simulations with 100~pc resolution. According to this comparison \revv{for the progenitors of Milky Way-sized galaxies probed by our simulations,} we expect that the stellar masses of the observed clumps are overestimated by as much as an order of magnitude, and that the sizes of these clumps are also overestimated by factor of several, due to a combination of spatial resolution and projection. The masses of young stars contributing most of the UV emission can also be overestimated by factor of a few. We find that most clumps of young stars present in a simulation at one time dissolve on a timescale shorter than $\sim$150~Myr. Some clumps with dense cores can last longer but eventually disperse. Most of the clumps are not bound structures, with virial parameter $\alpha_{\rm vir}>1$. We find similar results for clumps identified in mock maps of H$\alpha$ emission measure. We examine the predictions for effective clump sizes from the linear theory of gravitational perturbations and conclude that they are inconsistent with being formed by global disc instabilities. Instead, the observed clumps represent random projections of multiple \revv{compact} star-forming regions.
\end{abstract}


\section{Introduction}

While most nearby L* galaxies present disc-like morphologies, deep observations with the {\it Hubble Space Telescope} (HST) reveal irregular and clumpy shapes of high-redshift galaxies at rest-frame ultraviolet (UV) and optical wavelengths \citep{Elmegreen:2007aa, Overzier:2010aa, Swinbank:2010aa}. Integral field spectroscopic surveys of these galaxies show both rotation and turbulent motions \citep{Forster-Schreiber:2006aa, Shapiro:2008aa, Wisnioski:2015aa}. The fraction of galaxies that are clumpy at rest-frame UV evolves with time and varies with galaxy mass. The clumpy fraction for star-forming galaxies increases from $z\simeq8$ to $z\simeq1-3$, reaches a peak, and subsequently decreases until $z\simeq0$  \citep{Murata:2014aa, Guo:2015aa, Shibuya:2016aa}. This trend is similar to the evolution of the cosmic star formation rate (SFR) density \citep{Madau:2014aa}. Moreover, the clumpy fraction tends to increase with SFR at $z\simeq0-2$ \citep{Shibuya:2016aa}, suggesting a correlation with star formation activity. The clumpy fraction decreases with stellar mass in galaxies at $z\simeq 0.8-2$ \citep{Tadaki:2014aa, Guo:2015aa}.

Some of the earliest found clumpy galaxies were characterized as "chain", "tadpole", and "clump cluster" galaxies \citep{Elmegreen:2004ab,Elmegreen:2005aa}. There are typically 2-8 clumps per galaxy, with estimated stellar mass $\sim10^7-10^9\Msun$ \citep{Forster-Schreiber:2011aa, Guo:2012aa,Soto:2017aa}. The clumps typically have high SFR, resembling mini-starbursts in their galaxies \citep{Bournaud:2015aa,Zanella:2015aa}. While the clumps contribute only a few percent individually and $\sim$20\% altogether to their host galaxy's total luminosity, their contribution to SFR is larger, $\sim$10-50\% \citep{Forster-Schreiber:2011aa,Guo:2012aa,Wuyts:2012aa}. The SFR of individual clumps varies from $10^{-1}\Msun\,\mathrm{yr}^{-1}$ \citep{Soto:2017aa} to $1-10\Msun\,\mathrm{yr}^{-1}$ \citep{Guo:2012aa}. The inferred ages of the clump stars range from $10^6$ to $10^{10}$\,yr \citep{Soto:2017aa}. There is also wavelength dependence: clumps identified at different wavebands do not fully overlap \citep{Forster-Schreiber:2011aa}. 
 
The clumpy structure is also observed in H$\alpha$ \citep[e.g.][]{Livermore:2012aa,Mieda:2016aa} and CO maps \citep[e.g.][]{Swinbank:2010ab,Dessauges-Zavadsky:2017ab}. Kinematic studies of high-redshift massive clumpy galaxies ($M\sim10^{10.6}\Msun$) in H$\alpha$ emission using SINFONI/VLT show that these galaxies are turbulent and rotation dominated. Some galaxies have a massive stellar bulge \citep{Genzel:2008aa,Genzel:2011aa}.

Typical sizes of clumps in HST images $\sim$1\,kpc are at the limit of angular resolution at high redshift. Gravitational lensing has afforded us a magnified view of these galaxies. \citet{Adamo:2013aa} identified 30 clumps of $10^6-10^9\Msun$ in a lensed spiral galaxy Sp 1149 at redshift 1.5 with spatial resolution $\sim$100~pc. \citet{Girard:2018aa} observed a lensed rotating galaxy at $z=1.59$ and identified three H$\alpha$ clumps, which together contribute $\sim$40\% of total SFR inferred from the H$\alpha$ flux. The SFR density in these clumps is $\sim$100 times higher than in nearby HII regions. \citet{Livermore:2012aa} obtained the luminosity function of clumps with median source plane spatial resolution $\sim$360~pc and compared it with the luminosity function of ${\rm HII}$ regions in galaxies at $z\approx$1-1.5. They conclude that high-redshift clumps are ${\rm HII}$ regions that are larger and brighter than local ${\rm HII}$ regions. The clump sizes in lensed galaxies are smaller than those in unlensed galaxies. \citet{Jones:2010aa} found clumps with diameter 300~pc-1~kpc in lensed galaxies at $z$=1.7-3.1 with spatial resolution achieving $\sim$100~pc. \citet{Livermore:2015aa} extracted 50 star-forming H$\alpha$ and H$\beta$ clumps with sizes in the range 60~pc - 1~kpc in 17 lensed galaxies at 1<$z$<4. \citet{Wuyts:2014aa} found clumps of diameter $\sim$300-600~pc in a highly magnified lensed galaxy at $z=1.70$ and found a radial gradient of their rest-frame UV colour. \citet{Johnson:2017ab} found star-forming clumps of radius smaller than 100~pc in a lensed galaxy at $z=2.5$. \citet{Olmstead:2014aa} quantified relative stellar-to-nebular extinction in two $z$=0.91 galaxies with $\sim$0.3~kpc resolution. They found that the integrated extinction measurements agree with other studies in that the ionized gas is more obscured than stars. However, when examining on a clump-by-clump basis, they show that the hypothesis that stars and ionized gas experience identical extinction cannot be ruled out.

Unlike the clumpy structure observed in rest-frame UV, ALMA observations of cold dust of massive ($M_*\sim10^{11}\Msun$) star forming galaxies at $z\sim3$ with 200 pc resolution show smooth, disc-like morphology \citep{Rujopakarn:2019aa}.

The puzzling appearance of giant clumps has inspired many theoretical studies that have investigated the formation and evolution of these high-redshift clumps. They include isolated disk simulations \citep[e.g.][]{Tamburello:2015aa, Inoue:2018aa} and cosmological zoom-in simulations \citep[e.g.][]{Ceverino:2010aa,Oklopcic:2017aa}. The clumpy fraction in simulations decreases from high redshift ($z\approx2$) to low redshift \citep[e.g.][]{Buck:2017aa,Mandelker:2017aa}. These studies identify clumps using a variety of methods: in projected gas density maps  \citep[e.g.][]{Oklopcic:2017aa,Benincasa:2018aa},  projected stellar density maps \citep[e.g.][]{Mayer:2016aa}, mock observational maps \citep[e.g.][]{Hopkins:2012aa,Moody:2014aa,Buck:2017aa}, in 3D gas or stellar density distributions \citep[e.g.][]{Mandelker:2014aa,Mandelker:2017aa}, or as gravitationally bound objects \citep[e.g.][]{Tamburello:2015aa,Benincasa:2018aa}. Clumps identified in one type of maps do not necessarily correspond to clumps found in other maps \citep{Moody:2014aa}. In stellar maps, galaxies are only clumpy in UV light, but not in projected stellar mass density \citep{Buck:2017aa}. 

The origin of these high-redshift clumps is yet unclear. There are two general scenarios for clump formation: one in which clumps grow through gravitational instability within galactic discs, the other in which clumps are caused by external perturbations, such as mergers. The internal scenario, including violent disc instability and spiral arm instability, is supported by many simulations \citep[e.g.][]{Ceverino:2010aa, Genel:2012aa, Inoue:2018aa} and observations \citep[e.g.][]{Elmegreen:2007aa, Genzel:2008aa, Guo:2012aa, Zanella:2015aa}. Studies of nearby turbulent disc galaxies that resemble the high-redshift clumpy galaxies \citep{Fisher:2017ab} show that the clump sizes are consistent with the results of instabilities in self-gravitating gas-rich discs \citep{Fisher:2017aa}. If clumps form from fragmentation driven by turbulence, the clump stellar mass function should follow a power law of slope close to -2. \citet{Dessauges-Zavadsky:2018aa} derived the mass function of star-forming clumps at $z\sim1-3.5$ and found the power-law slope $\approx -1.7$ at $M_* > 2\times10^7\Msun$, in agreement with the turbulence-driven scenario. 

On the other hand, some simulation results indicate that {\it ex situ} mergers contribute to at least a portion of the clumps \citep{Mandelker:2017aa}, and that massive clumps could form from minor galactic mergers \citep{Mandelker:2014aa} and clump-clump mergers \citep{Tamburello:2015aa}. Observations of merging galaxies appear to support such merger-driven clump formation \citep{Puech:2009aa,Puech:2010aa,Guo:2015aa,Ribeiro:2017aa}.

The final fate of the clumps is also currently under debate. One alternative is that these clumps would migrate to the galactic center and potentially contribute to the galactic bulge \citep{Ceverino:2010aa, Inoue:2012aa, Perez:2013aa, Bournaud:2014aa}. Evidence for this scenario is built on the observed radial gradient of clump's colour or age, such that clumps closer to the galaxy center are older \citep{Noguchi:1999aa, Genzel:2008aa, Adamo:2013aa}. This scenario requires clump lifetime to be longer than a few orbital times. Another alternative is that these clumps dissolve in a relatively short time and may contribute to the thick disc, whereas the colour gradient could instead be a result of the inside-out disc growth \citep{Murray:2010aa, Genel:2012aa, Hopkins:2012aa, Buck:2017aa, Oklopcic:2017aa}. Some studies advocate both scenarios: low-mass clumps get disrupted in a short time, while more massive clumps survive and migrate to the center \citep{Genzel:2011aa, Mandelker:2017aa}. 

The interpretation of clump origins is complicated by possible overestimation of the clump masses and sizes due to limited angular resolution and sensitivity \citep[e.g.,][]{Tamburello:2015aa, Tamburello:2017aa, Cava:2018aa}. The observed kpc-scale clumps may also be clusters of clumps or blending of smaller structures \citep{Behrendt:2016aa}. Due to limited sensitivity, the observed clumps may be biased against low-mass structures   \citep{Dessauges-Zavadsky:2017aa}. These effects need to be taken into consideration when measuring clump properties.

In this work, we revisit the nature of giant clumps using our state-of-the-art simulations of galaxy formation. We explore how resolution and sensitivity affect the inferred clump mass and size, and investigate the final state of these clumps. Our high-resolution cosmological simulations include novel and most realistic modeling of star formation and stellar feedback, which allows us to produce galaxy structures that closely resemble observed high-redshift galaxies. In Section~\ref{sec:identification} we describe our simulations and clump identification method, both in 3D and in 2D projection. We present the properties of real 3D clumps and mock 2D clumps in our simulations in Section~\ref{sec:result}. In Section~\ref{sec:discuss} we discuss how simulated clumps compare with the observations and other simulations. We present our conclusions in Section~\ref{sec:conclude}. 

\section{Identification of clumps} \label{sec:identification} 

\subsection{Simulation suite}

We use a suite of cosmological simulations performed with the Adaptive Refinement Tree (ART) code \citep{Kravtsov:1997aa, Kravtsov:1999aa, Kravtsov:2003aa, Rudd:2008aa} and described in \citet{Li:2018aa} and \citet{Meng:2019aa}. All runs start with the same initial conditions in a periodic box of 4 comoving Mpc, so that the main halo has total mass $M_{200}\approx10^{12}\Msun$ at $z=0$, similar to that of the Milky Way. The ART code uses adaptive mesh refinement to increase spatial resolution in dense regions. There are $128^3$ root grid cells, setting the dark matter particle mass $m_{\rm DM}\approx10^6\Msun$. The finest refinement level is adjusted in runtime to keep the physical size of gas cells at that level between 3 and 6~pc. Because of strong stellar feedback, few cells reach the finest refinement level and the typical spatial resolution of molecular gas is 36--63~pc. \rev{We calculated the halo spin parameter $\lambda=J/(\sqrt{2}MVR)$ \citep{bullock_etal01} for our dark matter halos, where $J$ is the total angular momentum, $M$ is the halo mass, $V$ and $R$ are virial velocity and virial radius of the dark matter halo. The halo spin parameter varies in the range of $\sim$0.01-0.05, depending on the redshift. Here we list the halo spin parameters of our dark matter halos at the analyzed snapshots: $\lambda\sim$0.05 for SFE50 and SFE100 runs at $z$=1.50, $\lambda\sim$0.01 for SFE200 and SFE10 at $z$=1.78, and $\lambda\sim$0.04 for SFEturb run at $z$=1.98.}

The simulations include three-dimensional radiative transfer \citep{Gnedin:2001aa} of ionizing and UV radiation from stars \citep{Gnedin:2014aa} and extragalactic UV background \citep{Haardt:2001aa}, non-equilibrium chemical network of ionization states of hydrogen and helium, and phenomenological molecular hydrogen formation and destruction \citep{Gnedin:2011aa}. The simulations incorporate a subgrid-scale (SGS) model for unresolved gas turbulence \citep{Schmidt:2014aa,Semenov:2016aa}. Star formation is implemented with the continuous cluster formation (CCF) algorithm \citep{Li:2017ab} where each star particle represents a star cluster that forms at a local density peak and grows mass via accretion of gas until the feedback from its own young stars terminates the accretion. The feedback recipe includes early radiative and stellar wind feedback, as well as a supernova (SN) remnant feedback model \citep{Martizzi:2015aa, Semenov:2016aa}. The momentum feedback of the SN remnant model is boosted by a factor $f_{\rm boost}=5$ to compensate for numerical underestimation and to match the star formation history expected from abundance matching. To explore the variation of results with the speed of star formation, we ran a suite of simulations with different value of the local star formation efficiency (SFE) per free-fall time, $\epsff$. For full description of star formation and feedback recipe, see \citet{Li:2017ab, Li:2018aa}. 

Our simulation suite with $f_{\rm boost}=5$ produces the star formation history (SFH) expected for $10^{12}\Msun$ halos from abundance matching. The initial distribution of star clusters also matches observations of young clusters in nearby galaxies. The cluster initial mass function can be described by a Schechter function, the slope of which is close to the observed value -2 for $\epsff=0.5-1.0$. The fraction of clustered star formation correlates with the SFR density, consistent with observations for $\epsff=0.5-2.0$. The formation timescales of clusters in runs with $\epsff\geqslant0.1$ are shorter than 3~Myr, within the range of the observed age spread of young star clusters. 

In this paper we analyze several runs with different value of $\epsff$. The number after "SFE" in the run names corresponds to the percentage of local $\epsff$. In SFEturb run $\epsff$ is variable and turbulence-dependent (as implemented by \citealt{Semenov:2016aa}). The typical values are 3\%, with a log-normal scatter of about 0.3~dex \citep{Li:2018aa}. For each run we focus on the main galaxy in the simulation box. In all the runs the galaxies have a similar star formation history, reproducing the expectation of abundance matching. The galaxies also have similar axisymmetric stellar surface brightness profile \citep{Meng:2019aa}. The SFE has systematic effects on small spatial and temporal scales: higher SFE leads to shorter formation timescales for star clusters and more concentrated stellar feedback. This results in more bursty star formation rate and lower fraction of star-forming gas in galaxies with higher SFE. \rev{Our simulated galaxies have stellar masses $(2.4-8.2)\times10^9\Msun$ and neutral gas mass $(2.7-8.4)\times10^9\Msun$ at the analyzed snapshots ($z\approx1.5-3$, depending on the run). Other global properties of the simulated galaxies are listed in our previous paper \citep{Meng:2019aa}. }

\subsection{Identification criteria}

To match the procedures for finding clumps in rest-frame UV images of high-redshift galaxies, we made mock observations of our simulated galaxies and identified clumps in the mock images. We use the \textit{Flexible Stellar Population Synthesis} (FSPS) model \citep{Conroy:2009aa,Conroy:2010aa} to generate spectral energy distributions of all star particles and shift them to their corresponding redshifts in the simulation outputs.

To focus on rest-frame near-UV ($\sim$2800\AA) light, \citet{Guo:2015aa} detected clumps in HST/ACS F435W band for the galaxy redshift range $0.5\leqslant z < 1.0$, F606W band for $1.0\leqslant z < 2.0$, and F775W band for $2.0\leqslant z < 3.0$. \rev{Note that the CANDELS galaxies used in \citet{Guo:2018aa} have stellar masses $10^9-10^{11}\Msun$ at $z$=1.5-3, covering the range of our galaxy masses but extending to higher values.}We follow their choice and calculate the simulated galaxy fluxes in the latter two bands, because our outputs lie in the range $1 < z < 3$. First we bin the mock observational images in two-dimensional (2D) patches of 100~pc side (physical, not comoving) to avoid uneven resolution of cells at different refinement level. Then we degrade the images to patches of 500~pc to approximate the HST resolution at these redshifts. Galaxy orientation is chosen to be face-on, where the galaxy plane is determined using the shape tensor of neutral gas \citep{Meng:2019aa}.

We do not take into account the effect of dust because dust is not modeled explicitly in our simulations. Including dust attenuation could change the degree of clumpiness of the mock images, suppressing intrinsically bright clumps and enhancing less luminous ones \citep{Buck:2017aa}. 

We use the {\tt python} package {\tt astrodendro}\footnote{\url{http://dendrograms.org/}} to identify clumps in our mock observations. This package computes dendrograms, which are tree diagrams particularly useful for identifying hierarchical structures. A dendrogram contains \textit{leaves}, which have the highest values and no substructure, and combines them into  \textit{branches}, which merge hierarchically into the largest branch, the tree \textit{trunk}. We consider only leaves in the dendrogram of mock observational maps, since structures larger than leaves contain more than one clump. We focus our analysis on the brightest clumps, by setting a lower limit on the fractional contribution of each clump to the galaxy UV luminosity: $f_{\rm LUV} \equiv L_{\rm clump}^{\rm UV}/L_{\rm galaxy}^{\rm UV} > 3\%$, following \citet{Guo:2018aa}.

There are three parameters in the {\tt astrodendro} setup: {\tt min-value}, which is the minimum value of the surface brightness for a structure to be identified; {\tt min-delta}, which is the minimum difference between adjacent structures; and {\tt min-npix}, which is the minimum number of pixels required to form a structure. Experimentation showed that the results depend largely on {\tt min-npix}, but not the other two parameters. We set the value of {\tt min-npix} according to the expected observational resolution, but then vary this parameter to find the sensitivity of the results.

We set the parameter {\tt min-value} to be the typical surface brightness in the outer parts of the galaxy, about 26 mag~arcsec$^{-2}$, to include most of it in the dendrogram tree. Raising {\tt min-value} would make the smallest clumps disappear, most of which are excluded from our sample anyway due to the $f_{\rm LUV}$ cut. It would also make some of the largest clumps to be restricted only to their brightest parts.

The parameter {\tt min-delta} determines whether a peak is considered a clump or noise. We found {\tt min-delta} not as important as {\tt min-npix} in clump identification, as long as it is smaller than the mean difference among pixels that are in the region of clumps. We set the value of {\tt min-delta} at 0.5 mag~arcsec$^{-2}$.

\rev{Since high-redshift galaxies do not have well-defined thin discs, intrinsic clumps could be overlapping with each other, even when the galaxies are viewed face-on. Identifying clumps in 3D uncovers this projection effect. }We also identify clumps in three dimensions (3D), using dendrograms of stellar 3D density. We calculate the mass density of all stars and of stars younger than 100 Myr on a uniform 3D grid of cubes of 100 pc side. The $Z$-direction of this uniform 3D grid is perpendicular to the galaxy plane, so that a column of 3D cubes coincides with one 2D patch. This way we can associate each 2D clump with one or several 3D clumps in the same projected area. The 3D dendrogram builds a similar tree-like structure starting from regions of highest density. We identify 3D clumps using the density of young stars because young stars are the main contributor to UV luminosity, and the distribution of old stars does not appear clumpy. 

The choice of parameter {\tt min-value} and {\tt min-delta} for 3D clump identification is similar to that in 2D. We choose the parameter {\tt min-value} by examining the 3D light and young stellar density profile of the faintest clumps and take the value at the boundary where the profile dissolves into the background. {\tt min-delta} is set to be small enough that we do not artificially combine peaks due to this parameter. We take {\tt min-value}=$10^{-1.6}\Msun{\rm pc}^{-3}$ and {\tt min-delta}=0.2~dex for 3D clump identification.

\section{Results} \label{sec:result}

\subsection{Clump properties in simulations}

 \begin{figure*}
  \centering
  \includegraphics[width=0.86\textwidth]{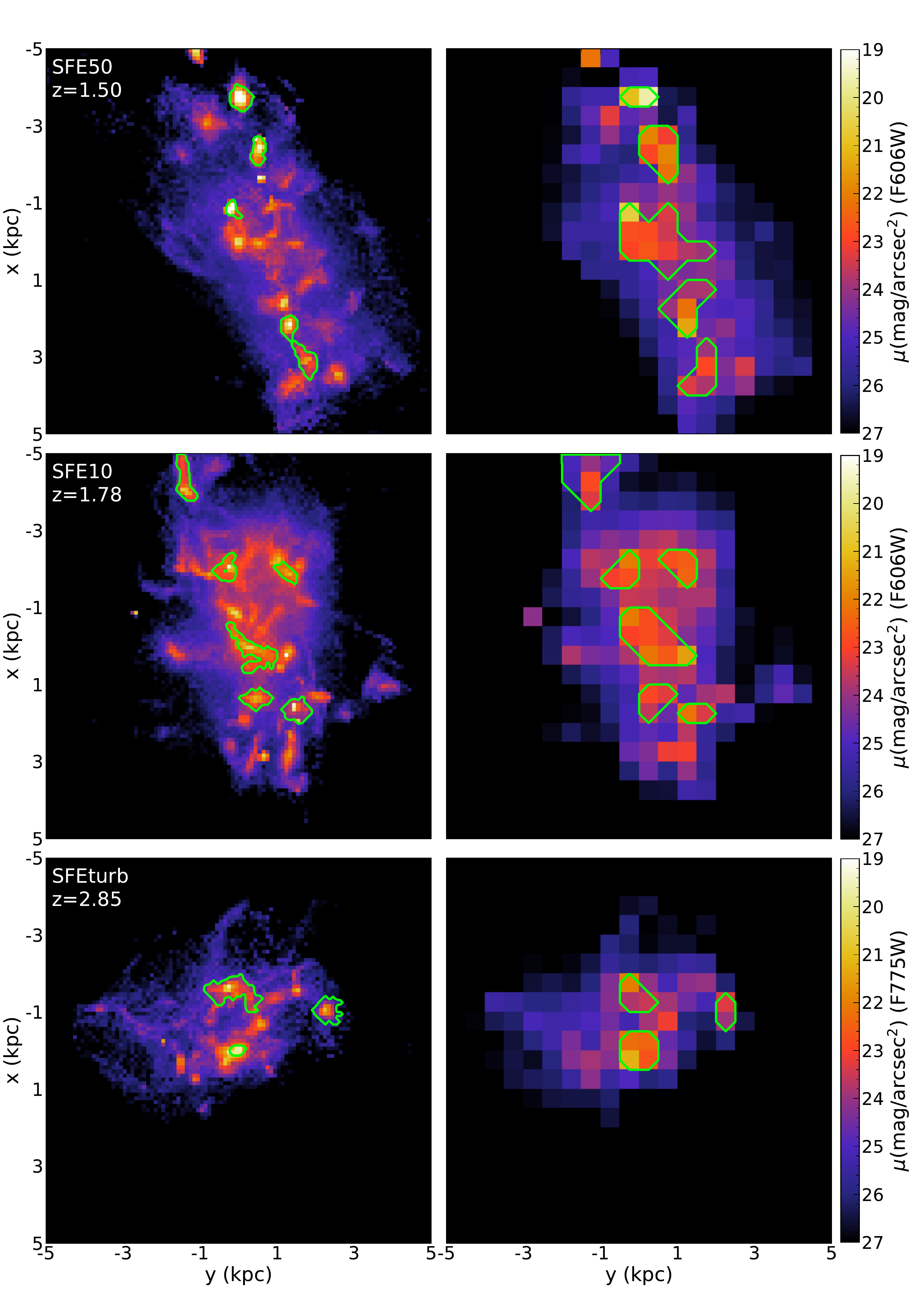}
\caption{Comparison of 2D and 3D clumps in mock images of simulated galaxies. Colour maps show the surface brightness in HST/ACS F606W band (except lower panels for SFEturb which are in the F775W band because of higher redshift of the simulation output) and do not include internal extinction. Left panels have pixels of 100~pc, right panels are degraded to HST resolution (500~pc) at these redshifts. Contours in the right panels show 2D clumps identified in the degraded maps using {\tt astrodendro} with {\tt min-npix}=2. Only clumps with fractional luminosity $>3\%$ are shown. Contours in the left panels show the most luminous 3D clump of young stars within the projected area of the corresponding 2D clump, with {\tt min-npix}=14.} \label{fig:summap}
 \end{figure*}
 
 \setcounter{figure}{0}
 \begin{figure*}
  \centering
  \includegraphics[width=0.86\textwidth]{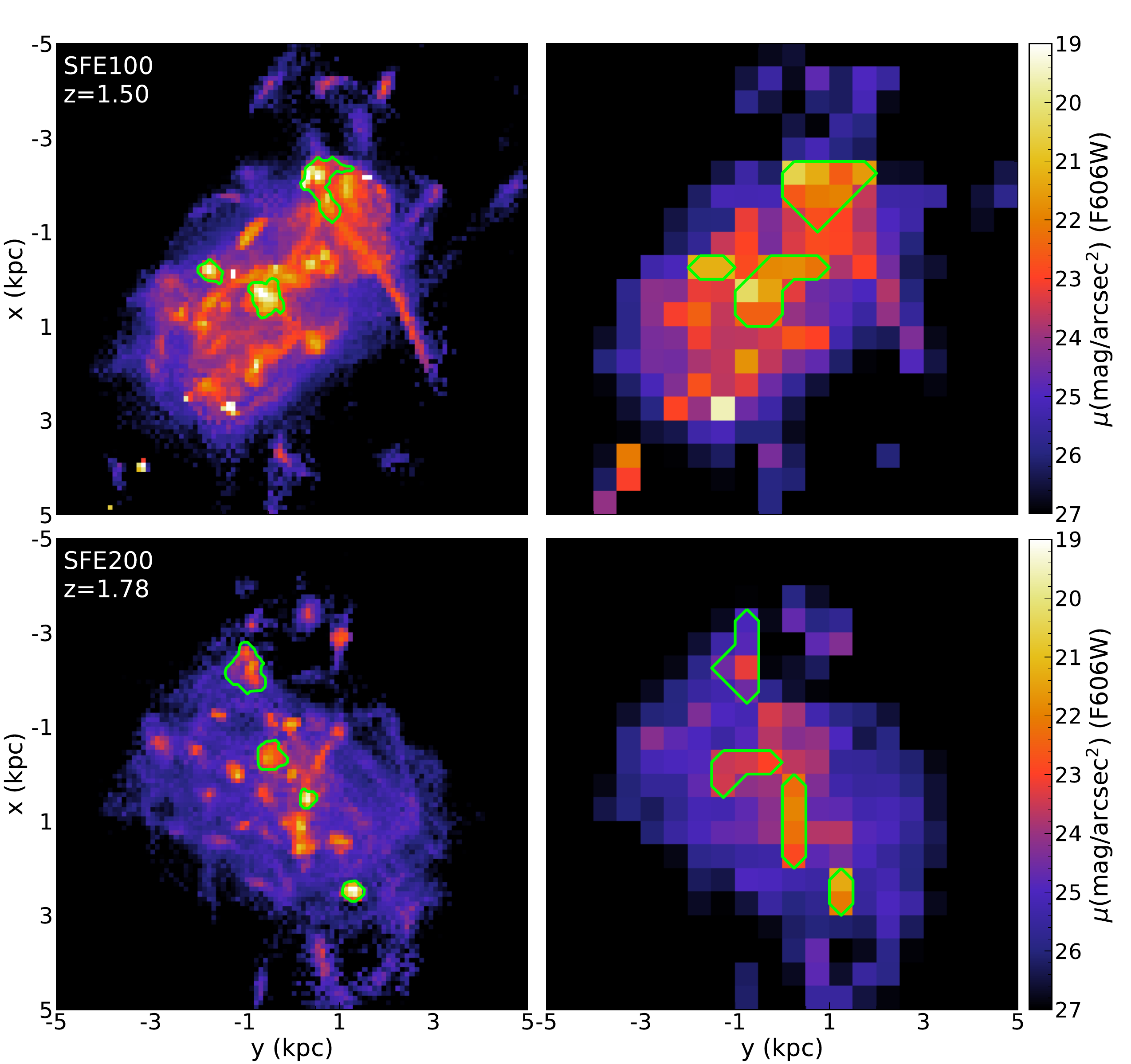}
  \caption{(continued)}
 \end{figure*}

Using the clump finding method described above, we identified clumps in 2D surface brightness maps and in 3D density grid of young stars. In a given snapshot there are several 2D clumps in the main galaxy, and 10--40 3D clumps, depending on the parameter {\tt min-npix}.

To explore the effects of parameter choice, we tried different values of {\tt min-npix} for the identification of 3D clumps. We aim to find separate 3D clumps of young stars that are large enough to account for most of the UV luminosity from the corresponding projected 2D clump. Our 3D patch size is fixed at 100~pc, so that larger {\tt min-npix} means larger minimum required clump volume. We concluded that {\tt min-npix}=14 is a good choice for 3D identification: lower value of {\tt min-npix} would only produce more small 3D clumps, while higher value of {\tt min-npix} would artificially lump together clumps that are already large enough. Also, if we use larger value of {\tt min-npix}, we would miss one very small and compact clump in SFE50 that accounts for 48\% of the rest-frame UV luminosity of its 2D counterpart. Based on these tests, we think {\tt min-npix}=14 gives the most stable result for 3D clump finding.

For 2D clump identification in degraded mock observations with patches of 500~pc, we take {\tt min-npix}=2, since it best matches the 2D clumps identified by eye. This choice also corresponds to the roughly kpc scale of the observed clumps.

Therefore, we adopt the values {\tt min-npix}=2 for 2D and {\tt min-npix}=14 for 3D clump finding. This choice implies a minimum size of the identified clumps. We define the effective clump size as the radius of a circle (in 2D) or a sphere (in 3D) with the same area/volume as the sum of patches in the clump, via {\tt Clump~Area}$\,\equiv \pi R_{\rm 2D}^2$ and {\tt Clump~Volume}$\,\equiv \frac{4\pi}{3}R_{\rm 3D}^3$. Our choice of {\tt min-npix} corresponds to the minimum effective radius of 400~pc in 2D and 150~pc in 3D. Note that \citet{Oklopcic:2017aa} uses similar minimum effective radius (125~pc) for 2D clump identification in FIRE simulations. 

\autoref{fig:summap} shows the identified 2D clumps and 3D clumps in five simulation runs with different SFE. Colour maps show the surface brightness of each galaxy in HST/ACS F606W (for $1\leq z <2$) and F775W (for $2\leq z<3$) bands. The spatial resolution for the left panels is a typical simulation patch of 100 pc, while the right panels are degraded to 500 pc to approximate the HST resolution at $z>1.5$. Contours in the right panels show 2D clumps identified in the degraded maps. For each of these clumps, in the corresponding left panel we show the projected contours of the 3D clump that contributes the most to the rest-frame UV luminosity within the area of that 2D clump. Reduced resolution tends to mix smaller clumps into one big peak, resulting in the kpc-scale clumps seen in the observations.

\begin{table*}
  \centering 
  \caption{Clump masses and sizes identified in 2D and 3D} \label{tab:clumps2D3D}
  \begin{threeparttable}
  \begin{tabular}{lcccccccccc}
   \toprule
    Run & $L_{\rm 2D}/L_{\rm gal}$\tnotex{tn:1} & Flux\tnotex{tn:2} & $M_{\rm 2D}$\tnotex{tn:3} & $R_{\rm 2D}$\tnotex{tn:4} & $M_{\rm 3D,tot}$\tnotex{tn:5} & $M_{\rm 3D,max}$\tnotex{tn:6} & $R_{\rm 3D}$\tnotex{tn:7} & $L_{\rm 3D,max}/L_{\rm 2D}$\tnotex{tn:8} & ${\rm SFR_{3D,max}}/{\rm SFR_{2D}}$\tnotex{tn:9} & $L_{\rm 3D,H\alpha}/L_{\rm 2D,H\alpha}$\tnotex{tn:10}\\ 
     &  & ($\mu$Jy) & ($10^7\Msun$) & (pc) & ($10^7\Msun$) & ($10^7\Msun$) & (pc) &  &  & \\ 
   \midrule
    SFE200  & 0.195 & 0.060 & 22.9 & 560 &  5.6 &  0.7 & 160 & 0.42 & 0.10 & 0.11\\
            & 0.176 & 0.054 &  4.9 & 400 &  1.1 &  1.1 & 200 & 0.95 & 0.97 &  -  \\
            & 0.074 & 0.023 & 15.4 & 560 &  2.1 &  0.8 & 230 & 0.37 & 0.56 & 0.88\\
            & 0.041 & 0.013 &  6.1 & 630 &  2.1 &  2.1 & 330 & 0.74 & 0.90 &  -  \\
   \midrule                                                     
    SFE100  & 0.198 & 0.213 & 34.9 & 800 &  9.8 &  6.4 & 300 & 0.57 & 0.52 & 0.43\\
            & 0.198 & 0.213 & 68.7 & 750 & 13.6 &  6.5 & 300 & 0.50 & 0.48 & 0.14\\
            & 0.073 & 0.078 & 12.4 & 400 &  1.3 &  1.3 & 220 & 0.46 & 0.65 &  -  \\
   \midrule
    SFE50   & 0.320 & 0.211 &  6.3 & 400 &  4.7 &  4.7 & 260 & 0.99 & 0.97 & 0.73\\
            & 0.201 & 0.132 & 58.2 & 940 &  6.8 &  0.7 & 150 & 0.48 & 0.09 & 0.82\\
            & 0.111 & 0.073 & 10.3 & 630 &  0.7 &  0.7 & 190 & 0.69 & 0.58 & 0.70\\
            & 0.090 & 0.059 & 24.3 & 630 &  3.9 &  0.8 & 210 & 0.51 & 0.17 & 0.19\\
            & 0.032 & 0.021 &  9.6 & 560 &  1.8 &  1.5 & 270 & 0.38 & 0.61 &  -  \\
   \midrule
    SFE10   & 0.236 & 0.108 & 82.9 & 800 &  8.3 &  3.7 & 280 & 0.21 & 0.41 & 0.26\\
            & 0.076 & 0.035 & 21.6 & 490 &  3.9 &  3.1 & 220 & 0.52 & 0.57 &  -  \\
            & 0.074 & 0.034 & 26.0 & 490 &  4.2 &  1.1 & 180 & 0.23 & 0.16 &  -  \\
            & 0.047 & 0.022 &  5.0 & 400 &  1.0 &  1.0 & 220 & 0.89 & 0.81 &  -  \\
            & 0.046 & 0.021 &  4.5 & 690 &  0.8 &  0.8 & 260 & 0.65 & 0.60 & 0.49\\
            & 0.042 & 0.019 & 14.6 & 490 &  1.2 &  1.2 & 230 & 0.56 & 0.83 & 0.35\\
   \midrule
    SFEturb & 0.372 & 0.091 & 35.6 & 560 & 20.3 &  8.0 & 160 & 0.31 & 0.29 & 0.28\\
            & 0.132 & 0.032 & 15.6 & 490 & 11.7 &  9.8 & 350 & 0.82 & 0.84 &  -  \\
            & 0.043 & 0.011 &  4.9 & 400 &  3.1 &  3.1 & 290 & 0.91 & 0.94 &  -  \\
   \bottomrule
  \end{tabular}
  \begin{tablenotes}
   \item[1] \label{tn:1} Fraction of the galaxy's luminosity contributed by the 2D clump. Only clumps with fractional luminosity $>3\%$ are listed. 
   \item[2] \label{tn:2} Clump's flux in HST/ACS F606W waveband, except for SFEturb run which is in F775W because of higher output redshift. 
   \item[3] \label{tn:3} Stellar mass of 2D clumps, defined using surface density as: $M_{\rm 2D}=\Sigma_*\times{\rm Clump Area}$. 
   \item[4] \label{tn:4} Effective radius of 2D clumps, defined as: Clump Area = $\pi R_{\rm 2D}^2$. 
   \item[5] \label{tn:5} Sum of stellar mass of all 3D clumps within the projected area of the 2D clump. 
   \item[6] \label{tn:6} Stellar mass of the most luminous 3D clump within the projected area of the 2D clump. 
   \item[7] \label{tn:7} Effective radius of 3D clumps, defined as: Clump Volume = $\frac{4\pi}{3}R_{\rm 3D}^3$. 
   \item[8] \label{tn:8} Ratio of luminosity of 3D clumps to luminosity of 2D clumps in each galaxy's corresponding waveband. 
   \item[9] \label{tn:9} Ratio of star formation rates in 3D clumps and 2D clumps, averaged over 100 Myr. 
   \item[10] \label{tn:10} Ratio of H$\alpha$ luminosity in 3D clumps and 2D clumps, for corresponding 2D UV clumps (matched by eye). "-" means there is no corresponding H$\alpha$ 2D clump for a 2D UV clump. 
  \end{tablenotes}
  \end{threeparttable}
 \end{table*}
 
The SFE50 run has five 2D clumps. The top one is the brightest while the middle one is the largest. The large clumps are massive and extended and most of them include multiple 3D clumps in projection. This leads to significant overestimation of both mass and size. The most luminous corresponding 3D clump is bright and compact, but there is also another 3D clump that accounts for 15\% of the 2D luminosity, and four other clumps, each contributing a few percent of the luminosity. At the top edge of the galaxy there is another structure which produces a single 2D pixel, but since we require at least two pixels for 2D clumps, this single pixel is not identified as a clump. 

The SFE10 run has a big 2D clump in the center, which is also a complex of multiple 3D clumps. Some of its smaller 3D clumps show elongated structure. The last snapshot of SFEturb run is at $z=2.85$, much earlier than the other runs, so the main galaxy in SFEturb run is smaller. It has only three 2D clumps; one of the corresponding 3D clumps is compact, and the other two are more extended. The SFE100 run has three 2D clumps, two of which are large and extended, and so are the corresponding 3D clumps. The SFE200 run has four 2D clumps, all of which look elongated to some extent. The corresponding 3D clumps are small and compact, with many young stars located between the clumps, further leading to overestimation of mass and size of the 2D clumps.

\autoref{tab:clumps2D3D} lists the properties of 2D and 3D clumps. The masses of the 2D clumps are in the range $10^7-10^9\Msun$, in agreement with clump masses in \citet{Guo:2018aa}. While the effective sizes of 2D clumps are of order kpc, as in the observations \citep{Forster-Schreiber:2011aa,Guo:2012aa,Soto:2017aa}, the sizes of 3D clumps are significantly smaller. The projected area of some 2D clumps may contain multiple 3D clumps. To demonstrate the effects of mixing and projection, we include two estimates of 3D mass: the total mass of all 3D clumps that contribute more than 32 magnitude in the corresponding HST band (but not necessarily centered) within the 2D clump column, and the mass of only one clump that contributes the highest rest-frame UV luminosity. Although mixing of multiple 3D clumps increases the mass by a factor of a few in some cases, inclusion of stars that are not in any identified 3D clump contributes much more to the overestimation of 2D clump masses.

The flux of the 2D clumps in their corresponding wavebands is similar to that in the observations of \citet{Guo:2018aa}. The flux of their clumps at $1.5<z<3$ in F606W and F775W bands is mostly between 0.02 to 0.3 $\mu$Jy. Similarly, the largest of our simulated 2D clumps can contain between 20\% and 37\% of their host galaxy UV luminosity.

In the remaining columns we show the ratio of luminosities and SFRs of the most luminous 3D clump and the corresponding 2D clump. Although the masses of 2D clumps are significantly overestimated, the difference in luminosity is smaller. This is because we identify 3D clumps using young stars and they are the main contributor to the rest-frame UV luminosity. Similar to the luminosity ratio, the SFR in 2D clumps is also less overestimated than mass and size, because UV luminosity traces young stars.

\rev{The total contribution of 3D clumps to SFR in 100~Myr is 50-80\%. Each individual 3D clump typically contributes a few percent to total SFR, up to $\sim$30\%.}The contribution of individual 2D clumps to the total SFR typically varies from a few percent to $\sim$15\%, and can be up to 60\%. This is generally consistent with the observational results of \citet{Guo:2015aa} who find that the clumps typically contain a few percent of the total SFR, and of \citet{Soto:2017aa} who find median clump contribution of 5\% to the total SFR.

\begin{figure*}
  \includegraphics[width=\textwidth]{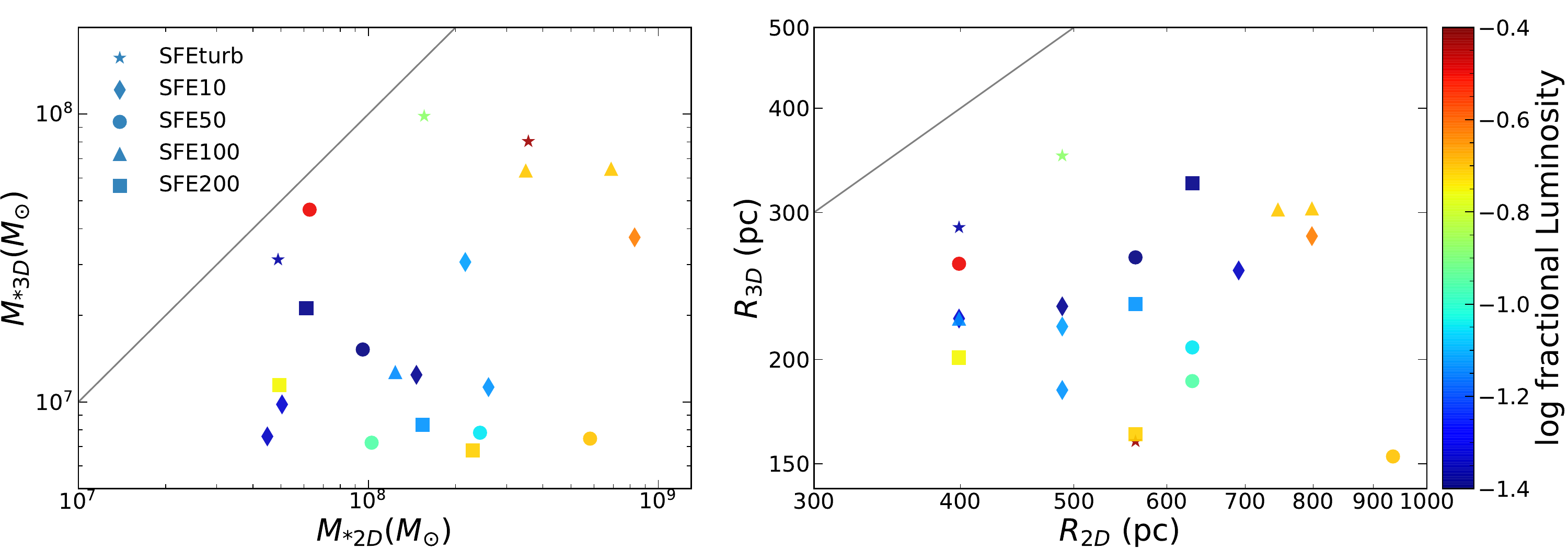}
  \vspace{-4mm}
\caption{Stellar mass (left panel) and effective radius (right panel) of clumps identified in 2D degraded mock observation map (resolution 500~pc) with {\tt min-npix}=2 and 3D density of young stars (younger than 100 Myr) identified with {\tt min-npix}=14. Here stellar mass in 2D is total stellar mass within the projected column, and stellar mass in 3D is total stellar mass of the most luminous 3D clump within that 2D clump. Effective radius is the radius of a circle (in 2D) or a sphere (in 3D) that has the same area/volume as the corresponding clump. The colour of each point is the fractional luminosity of the 2D clump, $L_{\rm 2D}/L_{\rm gal}$. Gray thick solid lines are 1:1.}
 \label{fig:2D3D}
\end{figure*}

\subsection{Comparison of 2D and 3D masses and sizes}

The first obvious result of our comparison of clumps identified in 2D and 3D is the differences in mass and size. We compare the properties of 2D and 3D clumps in \autoref{fig:2D3D}. Each point corresponds to the 3D clump that contributes the most to the luminosity of the 2D clump when projected onto the galaxy plane. The stellar mass of the identified 2D clumps range from several times $10^7\Msun$ to $10^9\Msun$, while few of the corresponding 3D clump masses exceed $10^8\Msun$. All the points are below the 1:1 line, which means that 2D clump masses and sizes are overestimated.

The overestimation of the 2D clump mass can be as large as an order of magnitude. Fractional luminosity of 2D clumps is not predictive of the extent of mass overestimation. The most overestimated 2D clumps are also more likely to contain multiple 3D clumps of comparable mass. Combining multiple snapshots, the most massive clump mass in both 2D and 3D seems to increase with decreasing SFE. The most overestimated clumps tend to have large fractional luminosity, since they are more likely to be combination of multiple clumps. 

Similarly, the effective sizes of 2D clumps are also overestimated, especially for those that are made up of multiple 3D clumps. The ratio of 2D to 3D effective radii can be a factor of several. We do not see any correlation between clump sizes and SFE. 

All of the 3D clumps have intrinsic sizes below the HST resolution of $\sim500$~pc, even after combining simulation grid cells to uniform patches of 100~pc. Therefore, the sizes of clumps are \textit{always overestimated} in observations, unless they benefit from the magnification by gravitational lensing.

\subsection{Clump longevity}

Important questions for the interpretation of these giant clumps are: Are they gravitationally bound? How long do they remain identifiable as distinct clumps? One way to address this is to look at how the average distance between pairs of clump stars varies over time. If majority of stars are self-bound, the average distance should not change much. On the other hand, if the stars that appear to be in a clump are unbound after the initial gas clearance, then their average distance would increase steadily over time. The maximum rate of expansion is set by the velocity dispersion of all stars in that region of the galaxy. Below we conduct several tests of the boundedness and longevity of the clumps in our simulations.

\begin{figure}
  \includegraphics[width=\columnwidth]{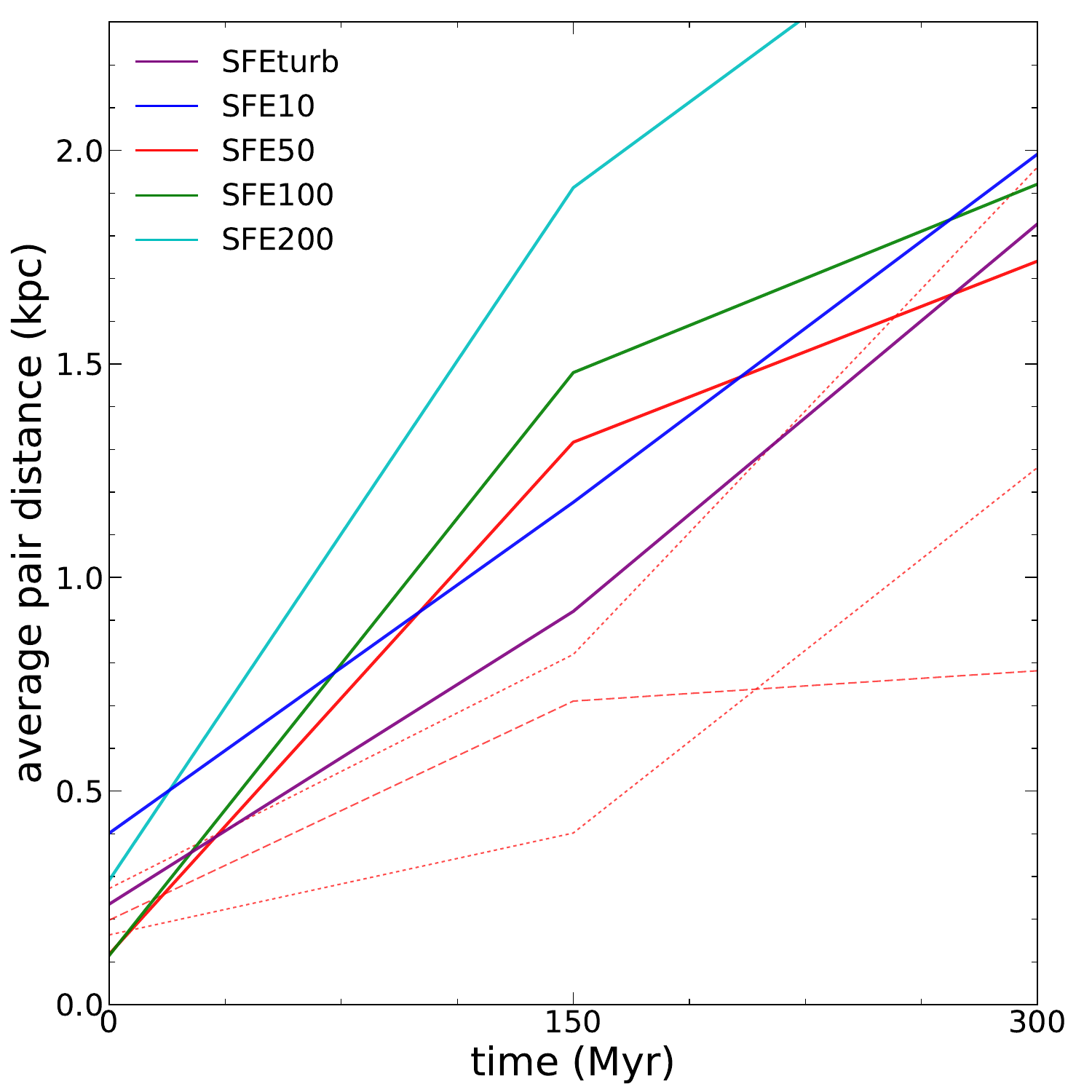}
\vspace{-5mm}
\caption{Evolution of mass-weighted average pair distance of young stars in 3D clumps over three consecutive snapshots. The time interval between snapshots is about 150~Myr. Solid lines show the most massive clump in each run. For SFE50 we show also the second most massive clump (dashed line) and the less massive ones (dotted lines). The upper limit of the y-axis corresponds to the distance traveled at 15~km/s over 150~Myr.}   \label{fig:disttime}
\end{figure}

\autoref{fig:disttime} shows the evolution of the average distance between pairs of young stars in 3D clumps, over the last three available simulation snapshots. After identifying a clump in the first snapshot, we calculate the mass-weighted average pair distance of the same stars in the following two snapshots, about 150~Myr and 300~Myr later. This can be used as a proxy for characteristic size of the clump. We can see that the average sizes of most clumps increase significantly after only 150~Myr, which means they dissolve and spread out. There are some less massive clumps that spread out slowly after 150~Myr, but they eventually dissolve after 300~Myr. We note that this analysis is limited by the time between saved simulation snapshots, and that clumps could dissolve in shorter than 150~Myr. 
 
To estimate whether the 3D clumps are bound, we calculate the virial parameter 
 \begin{equation}
  \alpha_{\rm vir} = a\frac{\sigma^2R}{GM}, \label{eq:alphavir}
 \end{equation}
where $M$ is clump mass, $R$ is the effective radius, and $\sigma$ is the 3D velocity dispersion of young stars. We take the parameter $a=5/3$ for a constant density sphere \citep{Bertoldi:1992aa}. Values of $\alpha_{\rm vir}\lesssim 1$ would indicate the clump is bound, although the precise value depends on the structure of the stellar distribution.

To quantify the evolution of the mass and size of a clump of young stars identified in one simulation snapshot, we look for clumps in the same group of stars in the next two consecutive snapshots. We keep the same density threshold for clump identification, so that stellar particles that dissolve into the background are not counted in the successor clumps. Then we calculate the fraction of mass of the original stars that remain in the most massive successor clump. We also take the mass-weighted average distance between pairs of the original stars remaining in the successor clump as a measure of size of the successor clump. 

We find a weak correlation between $\alpha_{\rm vir}$ and the mass fraction of young stars that can still be found in the successor clump: clumps with high remaining mass fraction have small $\alpha_{\rm vir}\lesssim 1$. For example, all (20) but two clumps with remaining mass fraction above 70\% have $\alpha_{\rm vir}<1$. Equivalently, clumps with small $\alpha_{\rm vir}$ tend to retain more of their original mass: the median remaining mass fraction for clumps with $\alpha_{\rm vir}<1$ is 61\%, while the median for clumps with $\alpha_{\rm vir}>1$ is only 7\%. \citet{Mandelker:2017aa} similarly found that clumps with shorter lifetime have larger $\alpha_{\rm vir}$. 

\begin{figure}
  \includegraphics[width=\columnwidth]{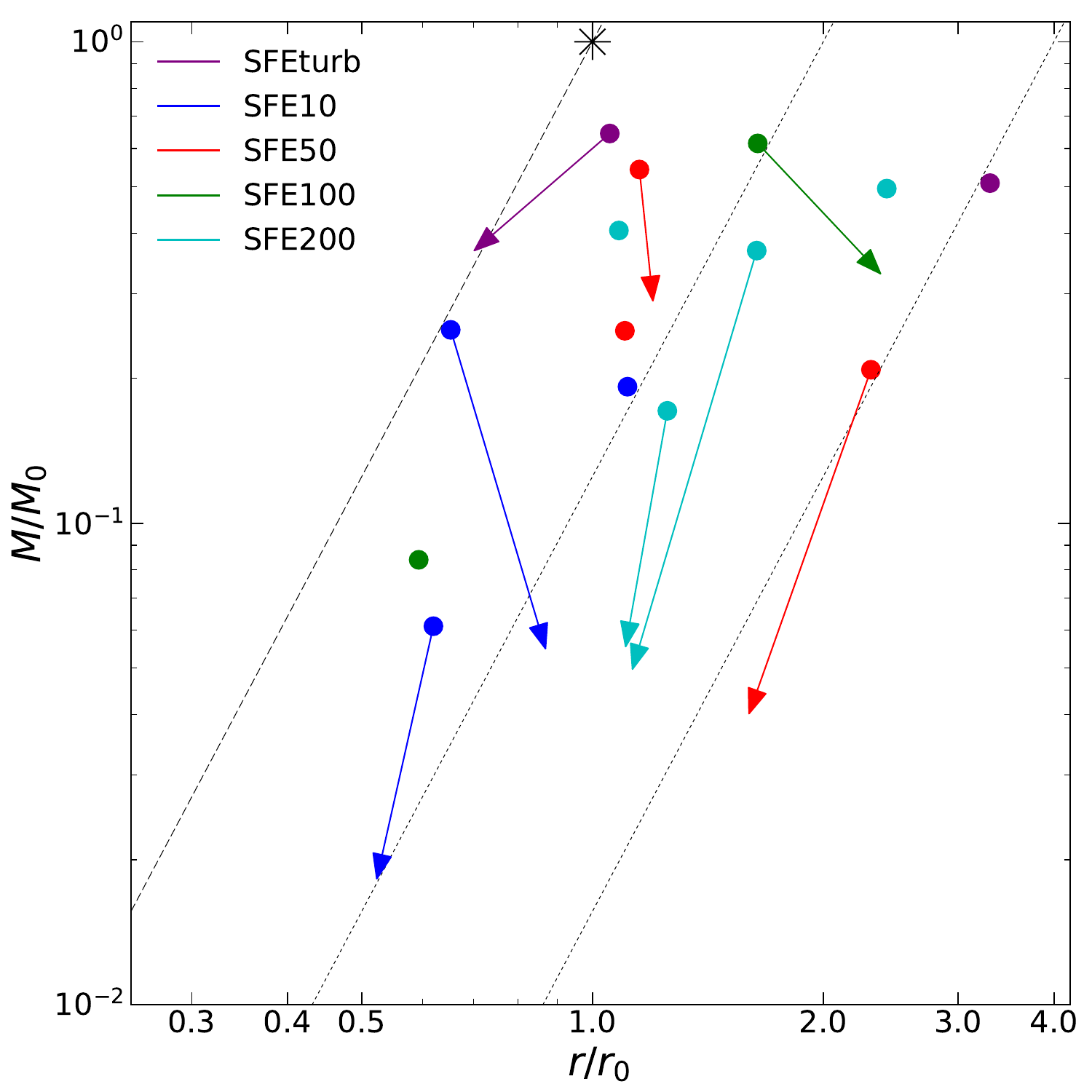}
\vspace{-5mm}
\caption{Evolution of the fraction of mass and mean separation of stars that remain in the most massive successor of clumps identified two snapshots before last available. The separation $r$ is calculated as the mass-weighted average distance between all pairs of stars. By construction every clump starts at (1,1), marked by the starry symbol. Circles show the ratios one snapshot later ($\sim$150~Myr), and arrows point to the values two snapshots later ($\sim$300~Myr). The dashed line marks the initial average density of the clumps, while dotted lines mark constant density 8 and 64 times lower. We note that, since $r$ is a proxy for size, "density" here is also a proxy for the actual density.} \label{fig:raindrop}
\end{figure}

The overall evolution of clump mass and size, relative to the moment it was first identified, is shown in \autoref{fig:raindrop}. Every clump begins at point (1,1) and then appears as a circle one snapshot later, with an arrow pointing to the second snapshot later. If there is only a circle without an arrow, it means that the successor was not found after two snapshots. Since we use the average pair distance as a proxy for size of the surviving clump, we can illustrate decrease in the average density with lines of constant $M/r^3$. Most circles (one snapshot later) lie below the dashed line, which means that the remaining clumps are less dense than the original clumps. Many arrows point in the direction of decreasing density, indicating that the remaining part of clumps keeps dissolving over time. After 300~Myr the density of most clumps decreases by an order of magnitude, after which they would no longer be interpreted as clumps.

There is one clump in SFEturb run whose average density increases between the second and third snapshots, as the the mass continues to decrease. In this case the less bound outer part of the clump dissolves, leaving the remaining inner part more dense. At the time it was first identified, the clump was gravitationally bound, with the virial parameter $\alpha_{\rm vir}=0.37$. After 125~Myr, the inner part of this clump still remains bound with $\alpha_{\rm vir}=0.19$. However, after 250~Myr, half of this clump is dissolved and spread out all over the galaxy, while core now has $\alpha_{\rm vir}=4.9$, with a significant increase in the velocity dispersion. We can expect that this clump will dissolve soon afterwards. 
 
To examine one possible fate of the clumps, that they migrate to the galaxy center due to dynamical friction and contribute to the bulge \citep{Ceverino:2010aa}, we calculated the evolution of the average galactocentric distance of the young stars in clumps. We found no net inward or outward migration of clump stars, similar to the conclusion of \citet{Buck:2017aa}. Instead of massive bulges, our galaxies contain nuclear star clusters that are consistent with being formed by in-situ star formation \citep{Brown:2018aa}.

It has been argued that feedback recipe affects the longevity of clumps in simulations. Simulations with only thermal feedback from supernovae often produce long-lived clumps which then migrate to the galaxy center \citep[e.g.][]{Ceverino:2010aa,Ceverino:2012aa,Mandelker:2014aa}. In simulations with strong momentum feedback, clumps usually dissolve in a short time $\lesssim$100~Myr \citep[e.g.][]{Oklopcic:2017aa}. The results of \citet{Mandelker:2017aa} with radiation pressure feedback fall into an intermediate category, where massive clumps survive and low-mass clumps disrupt. Our simulation results are consistent with the strong feedback regime: most clumps dissolve in a short time and do not migrate to the center. As shown by \citet{Li:2018aa} this strong feedback is required to match the star formation history of Milky Way-sized galaxies inferred from abundance matching, and therefore, we favour the conclusions obtained with our simulations.

\rev{We have available one weaker feedback run, which produced too high star formation rate and metallicity. The distribution of young stars in that run is smoother and the disc appears regular even at high redshift. If we run 3D clump identification algorithm over that galaxy, we get 3D clumps that are similar to clumps in other galaxies at face value, except that they are slightly larger. However, these 3D clumps would be less prominent than clumps in other galaxies, given the smoother distribution of young stars. Thus, these clumps should not be viewed as the same "clumps" expected in a clumpy galaxy. }

Some observed clumps appear to be old: e.g., \citet{Guo:2012aa} found clump ages in the range $10^{8}-10^{9}\,$yr, \citet{Soto:2017aa} in the range $10^{6}-10^{10}\,$yr, \citet{Zanella:2019aa} in the range $10^{6}-10^{9}\,$yr. The oldest end of these intervals is used to argue the longevity of the clumps. Some clumps even appear older than the underlying disc stellar population. However, inferred age can be severely affected by contamination by disc stars, as well as measurement systematics. When we calculate the mass-weighted average age of all stars in the clump regions in our simulations, we also find rather large ages \mbox{0.1-1~Gyr}, because of the large amount of old stars that do not contribute much to the rest-frame UV light. The true age of stars producing most of the UV light is of course below 100~Myr and cannot be used to set constraints on the dynamical longevity of the clumps.

\begin{figure*}
  \includegraphics[width=0.7\textwidth]{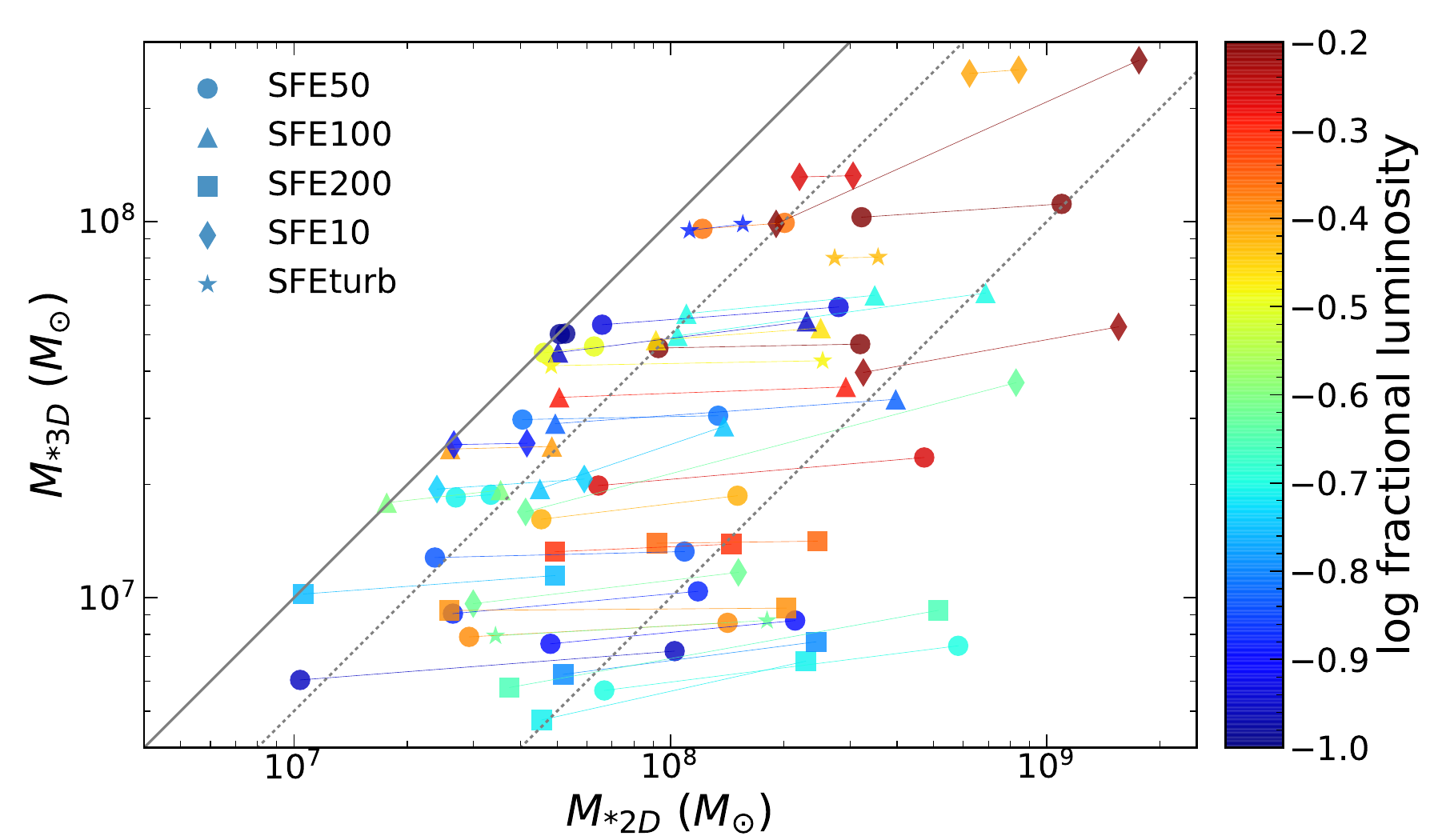}
  \vspace{-1mm}
\caption{Masses of 2D clumps and the most luminous 3D clumps within them in different snapshots at redshifts from 1.5 to 3 (between 2 and 8 snapshots per run).
Lines between symbols connect the mass of young stars (age less than 100~Myr) to the mass of all stars. To reduce the number of plotted points, we include only 2D clumps that contribute fractional luminosity $>10\%$. The solid line shows 1:1 ratio, and the dotted lines show 1:2 and 1:10 ratios.} \label{fig:young2D3D} 
\end{figure*}

\section{Discussion} \label{sec:discuss} 

\subsection{Effects of angular resolution}

In our simulations, masses of the 2D clumps are overestimated by about an order of magnitude. Comparison of clump masses in high-redshift field galaxies with those magnified by gravitational lensing points to a similar overestimation due to insufficient resolution. The observed clump stellar mass increases from around $10^7\Msun$ for lensed galaxies to $\sim10^9\Msun$ for field galaxies \citep{Dessauges-Zavadsky:2017aa, Cava:2018aa}. \rev{\citet{Cava:2018aa} observed the strongly lensed Cosmic Snake galaxy with the spatial resolution down to $\sim$30~pc, as well as the less strongly lensed Counterimage of the same galaxy with the resolution $\sim$300~pc. They found rest-frame UV clumps in the Counterimage with masses $\gtrsim10^8\Msun$, while the corresponding clumps in the Cosmic Snake image have masses down to $\sim10^7\Msun$. Our 3D clumps and 2D clumps show similar overestimation in mass.} Thus angular resolution is critical for correct characterization of giant clumps in high-redshift galaxies. 

Clump sizes in lensed galaxies range from several hundred pc to kpc \citep{Adamo:2013aa,Wuyts:2014aa,Livermore:2015aa}, smaller than clumps found in unlensed galaxies. \citet{Tamburello:2017aa} tested the resolution effect using the H$\alpha$ map smoothed with different Gaussian FWHM and found that clump sizes increase from $\sim$120~pc to $\sim$800-900~pc when FWHM increases from 100~pc to 1~kpc. 

Two effects can increase the observed mass of 2D clumps: mixing of multiple 3D clumps within the projected column and inclusion of stars that are not in clumps. From Table~\ref{tab:clumps2D3D} we can see that when clump masses are dramatically overestimated, the mass of intra-clump stars usually contributes more to the total mass than additional smaller 3D clumps. Including the mass of all 3D clumps usually increases the mass only by a factor of few, while including intra-clump stars can inflate the clump mass by over an order of magnitude. 

The amount of discrepancy between the 2D and 3D masses is much larger when counting stars of all ages, relative to counting only young stars. \autoref{fig:young2D3D} illustrates both ways of calculating the mass: lines connect the mass of young stars (lower left) to the mass of all stars (upper right) within the same clump. Counting all stars instead of only young stars is roughly equivalent to overestimating the SFR. From the plot we can see that SFR in 2D clumps could be overestimated by a factor of 3 or more. Similar result is found in \citet{Fisher:2017ab}, where blurred H$\alpha$ clumps in the DYNAMO-HST sample of low-redshift turbulent disc galaxies have $\sim 2-3$ times higher SFR than the associated full-resolution clumps. The overestimation of SFR is larger for more massive 2D clumps, because they may contain multiple distinct 3D clumps.

However, we do not find systematic dependence of the mass ratios on the output redshift. We plot masses of clumps identified at several epochs in all five runs: 8 outputs for SFE50 and 5 each for SFE100, SFE200, and SFE10 at redshift between 1.5 and 2, and 2 outputs for SFEturb at $z\approx3$. The distributions of points all these epochs are comparably broad.

A commonly used observational proxy for SFR is the luminosity $L_{\nu}$ in the wavelength range 1500-2800\AA\ \citep{Kennicutt:1998aa}: 
\begin{equation}
  {\rm SFR}\ (\Msun\,{\rm yr}^{-1}) = 1.4\times10^{-28}L_{\nu}\ ({\rm erg\,s^{-1}Hz^{-1}}). 
\end{equation}
This relation holds for the Salpeter stellar initial mass function (IMF). In this study we adopt the \citet{Kroupa:2001aa} IMF, which requires a correction factor of 0.64 \citep{Kennicutt:2012aa}. Using this relation, we compared the SFR calculated from the near-UV $L_{\nu}$ in our mock images to the actual SFR in the simulation, averaged over 100~Myr. The SFR inferred from $L_{\nu}$ is systematically 0.13~dex higher than the actual SFR, but with a large scatter of about 0.4~dex. The scatter mainly comes from the difference in age and metallicity of young stars. Stars older than 100~Myr could also contribute to the UV luminosity but are not counted in the SFR, which adds even more scatter. 

The ratio of the UV luminosities and SFRs of the 3D and 2D clumps are listed in \autoref{tab:clumps2D3D}. The difference in the two ratios is caused by the scatter in the SFR$-L_{\nu}$ relation discussed above. Most clumps have similar values of $L_{\rm 3D,max}/L_{\rm 2D}$ and ${\rm SFR_{3D,max}}/{\rm SFR_{2D}}$. Some clumps, however, show large discrepancies. For example, the first clump in SFE200 run and the second and fourth clumps in SFE50 run contain fairly large luminosity fractions, but smaller SFRs by a factor 3-5. This occurs because these 2D clumps combine multiple 3D clumps, and the selected most luminous 3D clump happens to be much younger than the other 3D clumps. For instance, in the second 2D clump in SFE50 run, the average age of the selected 3D clump is 5.9~Myr, while all other overlapping 3D clumps are about 50~Myr old. 

\subsection{Effects of numerical resolution}

We find that clumps identified at one snapshot in our simulations disperse and disappear in subsequent snapshots. To check that the dissolution of clumps is not caused artificially by the numerical relaxation effects, we calculated the half-mass relaxation time \citep{Spitzer:1969aa} for stellar particles making up the identified clumps:
 \begin{align}
  t_{\rm rh} &= \frac{0.17\,N}{\ln(\lambda N)}\sqrt{\frac{r_{\rm h}^3}{GM}} \nonumber \\ 
         &= \frac{0.78\, {\rm Gyr}}{\ln(\lambda N)}\frac{1\Msun}{m}\left(\frac{M}{10^5\Msun}\right)^{1/2}\left(\frac{r_{\rm h}}{1{\rm pc}}\right)^{3/2}.
 \end{align}
Here $m$, $M$, and $N$ are the average mass, total mass, and number of the stellar particles, respectively, and $r_h$ is the half-mass radius. We take $\lambda =0.2$ following \citet{binney_tremaine08}. Using another suggested value $\lambda =0.1$ does not change the results. 

The relaxation times for our clumps are typically several hundred Myr, with the smallest ones being about 100~Myr. Since it usually takes several relaxation times for a stellar system to dissolve, which is much longer than the time between two snapshots (about 150~Myr), we can conclude that the dissolution of the 3D clumps in our simulations is not due to numerical relaxation. 
 
\subsection{Gaseous clumps in line emission}

While we mainly focus on clumps of young stars in this paper, angular resolution also affects inferred sizes and masses of analogous clumps of ionized gas, found via their emission lines. For example, studies of H$\alpha$ emission from lensed high-redshift galaxies ($z=1-4$) find clumps with sizes ranging from $\sim$100~pc to 1~kpc \citep{Jones:2010aa,Livermore:2012aa,Livermore:2015aa}. The clump luminosity and SFR density increase with redshift. Measuring shifts of the H$\alpha$ provides also kinematic information on the high-redshift clumps \citep[e.g.,][]{Mieda:2016aa}.

H$\alpha$ emission traces even younger stars (age less than $\sim$10~Myr) than the UV light, because youngest stars contribute most of the ionizing radiation. To make more direct comparison with H$\alpha$ observations, we created mock maps of the H$\alpha$ emission measure. Similar to the UV clumps, we identify H$\alpha$ clumps in both 2D and 3D.

The emission measure is defined as $\int n^2\,ds$, where $n$ is the density of ionized hydrogen (here we use ionized hydrogen in cells with $T<20,000$~K, which contribute most to the cross-section), and $s$ is the length in the line of sight. In a simulation output the ionized hydrogen density is given on the adaptive mesh, with some cells on low refinement levels that are larger than our preferred uniform grid of 100~pc. Therefore, we resample all cells in the galaxy region to level 9, which corresponds to physical size \mbox{15-25}~pc, and then map the gas density on a uniform grid of 100~pc. We calculate the emission measure on this grid and then identify 3D H$\alpha$ clumps using {\tt astrodendro}. The parameters we use here are: {\tt min-pix}=14 and {\tt min-delta}=0.4~dex. The parameter {\tt min-value} is chosen using a similar approach of matching the background value as we did for 3D UV clumps. We choose different {\tt min-value} for different runs because these galaxies are at different redshift and have different $n_{\rm HII}$: {\tt min-value}=$\log n^2/{\rm cm}^{-6}$=1 for SFE50 and SFE100 runs, $\log n^2/{\rm cm}^{-6}$=0.2 for SFE10 run, $\log n^2/{\rm cm}^{-6}$=0.5 for SFE200, and $\log n^2/{\rm cm}^{-6}$=0 for SFEturb run.

To obtain a 2D map of the emission measure, we integrate it in the vertical direction and degrade it to 500~pc patches. Then we identify 2D H$\alpha$ clumps using the following {\tt astrodendro} parameters: {\tt min-npix}=2, {\tt min-delta}=0.1~dex, {\tt min-value}=200~${\rm pc\,cm^{-6}}$.

The H$\alpha$ clumps do not necessarily correspond to rest-frame UV clumps since they trace star formation on different timescales. For those UV and H$\alpha$ clumps that match spatially, we show the ratios of H$\alpha$ luminosity in 3D and 2D clumps in the last column of \autoref{tab:clumps2D3D}. For the UV clumps that do not have a corresponding H$\alpha$ clump, we leave dashes in the table. 

The first clump in SFE200 run has $L_{\rm 3D,H\alpha}/L_{\rm 2D,H\alpha}$ of only 11\%. The emission measure distribution in this galaxy is relatively smooth, thus producing this very small 3D H$\alpha$ clump.

We can compare the 3D/2D ratio of H$\alpha$ luminosities, which trace the SFR over $\sim$10~Myr, with the corresponding ratio of SFR averaged over 100~Myr. For many clumps they are close in value, and in these cases the 3D and 2D H$\alpha$ and UV clumps correspond to each other fairly well. Some clumps have large discrepancies, generally because the UV clump and the H$\alpha$ clump differ significantly in geometry. 

\rev{The H$\alpha$ light is more clumpy than UV light distribution, since it traces younger stellar population. The total fractional luminosity of 3D H$\alpha$ clumps ranges from $\sim$10\%-95\%, while the contribution to H$\alpha$ luminosity from individual 3D clumps ranges from less than 1\% to $\sim$90\%. This large range results from the short timescale of star formation traced by H$\alpha$ and thus the clumpiness of H$\alpha$ emission. Most individual 3D clumps contribute less than 10\% to total H$\alpha$ luminosity, while only one or two largest clumps contribute most of the H$\alpha$ flux. The gas in H$\alpha$ clumps is not bound, with virial parameters (\autoref{eq:alphavir})) larger than 1 by orders of magnitude.}

\subsection{Origins of giant stellar clumps}

Gravitational instability can induce fragmentation of the galactic disc, thus leading to growth of clumps. Stability criterion to linear axisymmetric perturbations for gaseous discs can be described by the Toomre parameter \citep{safronov60,toomre64}: 
\begin{equation}
  Q = \frac{\sigma\kappa}{\pi G\Sigma}, 
\end{equation}
where $\kappa$ is the epicycle frequency, $\sigma$ is the velocity dispersion, and $\Sigma$ is the mass surface density. A disk is unstable if $Q\lesssim1$. Observations \citep[e.g.][]{Puech:2010aa,Girard:2018aa} found that clumpy galaxies are marginally stable, with Toomre $Q\sim1$. Most H$\alpha$ clumps in high-redshift galaxies are located in regions where $Q$ is low \citep{Genzel:2011aa,Wisnioski:2012aa,Mieda:2016aa}. In galaxy formation simulations \citep{Inoue:2016aa, Oklopcic:2017aa} gas clumps also coincide with regions of $Q<1$, albeit with large scatter of $Q$ value. \citet{Inoue:2016aa} also found that stellar clumps coincide with regions of low stellar $Q$. 
 
In \citet{Meng:2019aa} we studied the distribution of $Q$ in our simulations, accounting for different phases of the gas and stars. Such multi-component $Q$ more accurately describes the linear stability criterion in realistic galaxies. We found that strong feedback from young stars disperses gas around them, leading to spatial \textit{anti-correlation} of dense gas and stars up to 50~Myr old. Here we analogously find that locations of 2D UV clumps do not coincide with the regions of low $Q$. This is because low $Q$ regions closely trace high gas density, while rest-frame UV clumps trace young stars. The average value of $Q$ in 2D UV clump regions ranges from 2.1 to 3.6, which is larger than the median value of $Q=0.5-1.0$ weighted by molecular gas mass in \citet{Meng:2019aa}. For H$\alpha$ clumps $Q=1.4-3.3$ is closer to the gas values, as expected for very young stars. The discrepancy in location of gas and young stars is also seen in the simulations of \citet{Oklopcic:2017aa}, where they find that the gas clumps coincide with the location of instantaneous star formation, but not of the SFR averaged over more than 10~Myr. 

If clumps form out of instability of a self-gravitating disc, the wavelength of the fastest growing perturbation is
\begin{equation}
 \lambda_T = \frac{2\sigma^2}{G\Sigma}.
\end{equation}
The effective radius of a fully-formed clump may not exactly correspond to this scale, but it can still be used as a rough guide. If we take $\lambda_T/2$ for $R_{\rm clump}$, and approximate the disc rotation velocity profile as flat, then $\kappa \approx \sqrt{2} V_{\rm rot}/R_{\rm disc}$, and we obtain
\begin{equation}
  \frac{R_{\rm clump}}{R_{\rm disc}} \approx \frac{\pi Q}{\sqrt{2}}\left(\frac{\sigma}{V_{\rm rot}}\right).
  \label{eq:rclump}
\end{equation}
Here $R_{\rm disc}$ is a characteristic size of the gas distribution. When the gaseous disc is marginally stable, i.e., $Q \sim 1$, we expect a linear relation between $R_{\rm clump}/R_{\rm disc}$ and $\sigma/V_{\rm rot}$. 

\citet{Fisher:2017aa} found the relation $R_{\rm clump}/R_{\rm disc}=(0.38\pm0.02)\,\sigma/V_{\rm rot}$ in nearby turbulent discs with properties closely resembling $z\sim2$ star-forming galaxies. They take twice the half-light radius of H$\alpha$ light as $R_{\rm disc}$ and identify clumps in H$\alpha+{\rm [N II]}$ map. They use flux-weighted $\sigma({\rm H_{\alpha}})$ and the modeled rotation velocity $V_{\rm rot}$ at 2.2 disc scalelengths. Note that the proportionality coefficient 0.38 is a factor of 6 smaller than that expected from the linear perturbation theory above. This already serves as a warning that the observed clump sizes may not be related to global disc instabilities.

We examined whether the 2D or 3D clumps in our simulations obey a similar relation. We take $R_{\rm disc}$ to be twice the half-light radius in our mock H$\alpha$ emission measure map. We use $\sigma_{\rm HII}$ in the brightest pixel and the circular velocity at galactocentric radius of 10 kpc as $\sigma$ and $V_{\rm rot}$. In the 25 snapshots of our 5 runs at different redshifts, most galaxies have the value of $\sigma/V_{\rm rot}$ between 0.2 and 1.2. The ratio of sizes $R_{\rm clump}/R_{\rm disc}$ for 2D clumps is in the range 0.1 to 0.4; for 3D clumps in the range 0.02 to 0.2. We do not find any correlation between the ratios $R_{\rm clump}/R_{\rm disc}$ and $\sigma/V_{\rm rot}$, either for 2D or 3D clumps. We also examined this relation for the projected UV clumps and did not find any correlation. This shows that our clump sizes are not set by gravitational instability in thin axisymmetric discs. In fact, if the relation (\ref{eq:rclump}) were to hold, we would expect clump sizes of the order $\lambda_T$, which is several kpc \citep{Meng:2019aa}, and therefore significantly larger than the effective radii of our identified clumps. \rev{We also calculated the radius $(M_f/\pi\Sigma)^{1/2}$ corresponding to the "fragmentation mass" $M_f=2\lambda_T\Sigma c_s/(\kappa f_g)$ defined by \citet{Tamburello:2015aa} through their equation 10, and found that it still overestimates the clump sizes, by factor of a few.}


We also investigated the dependence of clump properties on the specific SFR (sSFR). \citet{Shibuya:2016aa} found that the fraction of high-redshift clumpy galaxies increases with sSFR. \citet{Fisher:2017ab} found a positive relation between sSFR and the maximum fractional luminosity of low-redshift H$\alpha$ clumps, both blurred and with full resolution ($\sim$100~pc). Our H$\alpha$ clumps also show a positive correlation between the maximum fractional luminosity and sSFR. For most galaxies with sSFR$>2\times 10^{-10}\,{\rm yr}^{-1}$, the maximum fractional luminosity of 3D H$\alpha$ clumps is larger than 10\%, while the maximum fractional luminosity of 2D H$\alpha$ clumps (resembling their blurred clumps) is larger than 20\%. These results agree with those of \citet{Fisher:2017ab}. 

\subsection{Comparison with other simulation studies}

A number of studies have investigated the formation of giant clumps in galaxy formation simulations. Some simulation analyses identify 2D clumps in surface density maps. \citet{Oklopcic:2017aa} identified clumps using gas surface density in a massive galaxy in the FIRE simulation at $z=1-2$. They use a similar setup to our study: bin the surface density in patches of 50~pc and use {\tt astrodendro} with a minimum of 20 patches. Thus their clump sizes are several hundred pc, with the baryon masses $10^{6.5-9.5}\Msun$. \citet{Moody:2014aa} identified clumps in projected stellar mass maps and obtained clump masses $\sim10^{6.5}-10^9\Msun$. Our 2D clump masses are similar to the high-mass end of these clumps, but we miss the low-mass end possibly because of the adopted cut on fractional luminosity.

Other simulation studies identify clumps using 3D density, which is similar to our clump identification in 3D. \citet{Mandelker:2017aa} identified clumps in the 3D distribution of both the cold gas component and stellar component in the VELA simulation galaxies, with halo masses $10^{11}-10^{12}\Msun$ at $z=2$. They identified clumps as connected regions, which is similar to our hierarchical tree construction, containing at least 8 cells of $(70~{\rm pc})^3$ each and found clumps of baryon mass $\sim10^7-10^9\Msun$. \citet{Tamburello:2015aa} and \citet{Mayer:2016aa} used the {\tt SKID} algorithm to identify bound structures in their simulations. Their typical clump gas masses and stellar masses are $\sim10^7-10^8\Msun$. Our 3D clumps are similar to these clumps in mass. 

Some simulations investigate the effect of spatial resolution on identified clump masses using maps smoothed to different resolution. \citet{Tamburello:2017aa} identified clumps in mock H$\alpha$ images of their simulations, convolved with the 2D Gaussian aperture with full width at half maximum FWHM = 1~kpc and 100~pc. They found clump gas mass to be $\sim10^8-10^9\Msun$ for FWHM=1~kpc, and $\sim10^{6.5}-10^{8.5}\Msun$ for FWHM=100~pc. The clumps in the full-resolution H$\alpha$ maps are similar to the 3D bound structures found in \citet{Tamburello:2015aa}, but clumps found with FWHM=1~kpc have masses overestimated by an order of magnitude. Our results support this order-of-magnitude overestimation. We also agree on the amount of overestimate of the clump size: \citet{Tamburello:2017aa} find a median intrinsic radius $\sim$120~pc vs. $\sim$800~pc in the blurred images, while our 3D and 2D median clump sizes are $\sim$200~pc and $\sim$630~pc, respectively.
How much clump sizes are overestimated depends largely on the spatial resolution, and this comparison shows that the clump sizes in observations of unlensed galaxies are likely severely overestimated. 
 
\citet{Behrendt:2016aa} also identified clumps in mock H$\alpha$ maps. They convolved the surface density map with a 2D Gaussian of FWHM=1.6~kpc to mimic the instrumental response, and found clump baryon masses to be $(1.5-3)\times10^9\Msun$. \citet{Buck:2017aa} identified clumps in luminosity maps in the NIHAO galaxy sample, using both intrinsic clumps in non-dust-attenuated rest-frame U band and clumps in HST bands with dust taken into account. Their clump masses range from a few times $10^6\Msun$ to $10^9\Msun$, and sizes are $\sim300-900{\rm pc}$. The high-mass end is similar to our 2D rest-frame UV clumps in mock observation maps. \citet{Benincasa:2018aa} identified clumps in isolated disc galaxy simulations both in 3D using {\tt SKID} and in 2D gas surface density map using {\tt astrodendro}. They used 2D resolution of 10~pc and 100~pc, and found only a factor of a few difference between the 2D and 3D clump masses. This result guarantees that our 3D clumps identified on grid of 100~pc are not much overestimated compared to the intrinsic clumps. 

\section{Conclusions} \label{sec:conclude}

We investigated the nature of giant kpc-scale clumps in \revv{high-redshift progenitors of the Milky Way-sized galaxies}. We identified both 2D clumps in rest-frame UV mock HST observation maps with 500~pc resolution and intrinsic clumps in 3D density of young stars with 100~pc resolution (\autoref{fig:summap}). The 2D clumps are chosen to resemble observed giant clumps. The main results of our comparison are summarized below: 
\begin{itemize}
\item The masses and sizes of 2D clumps are overestimated due to limited angular resolution and projection of several 3D clumps along the line of sight. The overestimate of mass can be as large as an order of magnitude, and the overestimate of size can be a factor 2--3, compared to the most luminous corresponding 3D clump (\autoref{fig:2D3D} and \autoref{tab:clumps2D3D}). The intrinsic sizes of clumps (150--300~pc) are below the HST resolution at $z>1$, unless the source galaxy is strongly lensed.

\item Most clumps of young stars in our simulated galaxies dissolve on a timescale shorter than $\sim$150~Myr. The average pair distance between young stars in a clump  increases dramatically after 150~Myr (\autoref{fig:disttime}), and the density of the remaining parts of these clumps continues to decrease over time (\autoref{fig:raindrop}).

\item Most of the 3D clumps are not gravitationally bound structures, with the virial parameter $\alpha_{\rm vir}>1$. However, a few clumps with $\alpha_{\rm vir}<1$ are more likely to have a fraction of mass remain bound in the next simulation snapshot, after $\sim$150~Myr. 

\item Although total stellar masses of 2D clumps are significantly overestimated, the masses of young stars are only overestimated by factor of a few relative to the most luminous corresponding 3D clump (\autoref{fig:young2D3D}).

\item We created mock images of H$\alpha$ emission measure and compared the sizes of H$\alpha$ clumps with the expectation of the linear theory of global disc instabilities. We do not find the expected correlation between $R_{\rm clump}/R_{\rm disc}$ and $\sigma/V_{\rm rot}$, primarily because the clump sizes are much smaller than expected. Therefore, we conclude that the observed clumps are not the result of gravitational instabilities in thin axisymmetric discs, \revv{at least in Milky Way-sized galaxies}.
\end{itemize}

\section*{Acknowledgements}

We are grateful to Hui Li for producing the simulation suite used in this analysis. We thank the referee for useful comments. This work was supported in part by the National Science Foundation through grants 1412144 and 1909063.

\bibliographystyle{mnras}
\bibliography{lqbz,bzd,gc}

\begin{thebibliography}{}
\makeatletter
\relax
\def\mn@urlcharsother{\let\do\@makeother \do\$\do\&\do\#\do\^\do\_\do\%\do\~}
\def\mn@doi{\begingroup\mn@urlcharsother \@ifnextchar [ {\mn@doi@}
  {\mn@doi@[]}}
\def\mn@doi@[#1]#2{\def\@tempa{#1}\ifx\@tempa\@empty \href
  {http://dx.doi.org/#2} {doi:#2}\else \href {http://dx.doi.org/#2} {#1}\fi
  \endgroup}
\def\mn@eprint#1#2{\mn@eprint@#1:#2::\@nil}
\def\mn@eprint@arXiv#1{\href {http://arxiv.org/abs/#1} {{\tt arXiv:#1}}}
\def\mn@eprint@dblp#1{\href {http://dblp.uni-trier.de/rec/bibtex/#1.xml}
  {dblp:#1}}
\def\mn@eprint@#1:#2:#3:#4\@nil{\def\@tempa {#1}\def\@tempb {#2}\def\@tempc
  {#3}\ifx \@tempc \@empty \let \@tempc \@tempb \let \@tempb \@tempa \fi \ifx
  \@tempb \@empty \def\@tempb {arXiv}\fi \@ifundefined
  {mn@eprint@\@tempb}{\@tempb:\@tempc}{\expandafter \expandafter \csname
  mn@eprint@\@tempb\endcsname \expandafter{\@tempc}}}

\bibitem[\protect\citeauthoryear{{Adamo}, {{\"O}stlin}, {Bastian},
  {Zackrisson}, {Livermore}  \& {Guaita}}{{Adamo} et~al.}{2013}]{Adamo:2013aa}
{Adamo} A.,  {{\"O}stlin} G.,  {Bastian} N.,  {Zackrisson} E.,  {Livermore}
  R.~C.,   {Guaita} L.,  2013, \mn@doi [\apj] {10.1088/0004-637X/766/2/105},
  \href {http://adsabs.harvard.edu/abs/2013ApJ...766..105A} {766, 105}

\bibitem[\protect\citeauthoryear{{Behrendt}, {Burkert}  \&
  {Schartmann}}{{Behrendt} et~al.}{2016}]{Behrendt:2016aa}
{Behrendt} M.,  {Burkert} A.,   {Schartmann} M.,  2016, \mn@doi [\apjl]
  {10.3847/2041-8205/819/1/L2}, \href
  {http://adsabs.harvard.edu/abs/2016ApJ...819L...2B} {819, L2}

\bibitem[\protect\citeauthoryear{{Benincasa}, {Wadsley}, {Couchman}, {Pettitt}
  \& {Tasker}}{{Benincasa} et~al.}{2019}]{Benincasa:2018aa}
{Benincasa} S.~M.,  {Wadsley} J.~W.,  {Couchman} H.~M.~P.,  {Pettitt} A.~R.,
  {Tasker} E.~J.,  2019, \mn@doi [\mnras] {10.1093/mnras/stz1152}, \href
  {https://ui.adsabs.harvard.edu/abs/2019MNRAS.486.5022B} {486, 5022}

\bibitem[\protect\citeauthoryear{{Bertoldi} \& {McKee}}{{Bertoldi} \&
  {McKee}}{1992}]{Bertoldi:1992aa}
{Bertoldi} F.,  {McKee} C.~F.,  1992, \mn@doi [\apj] {10.1086/171638}, \href
  {https://ui.adsabs.harvard.edu/abs/1992ApJ...395..140B} {395, 140}

\bibitem[\protect\citeauthoryear{{Binney} \& {Tremaine}}{{Binney} \&
  {Tremaine}}{2008}]{binney_tremaine08}
{Binney} J.,  {Tremaine} S.,  2008, {Galactic Dynamics}.
Princeton, NJ: Princeton Univ. Press

\bibitem[\protect\citeauthoryear{{Bournaud} et~al.,}{{Bournaud}
  et~al.}{2014}]{Bournaud:2014aa}
{Bournaud} F.,  et~al., 2014, \mn@doi [\apj] {10.1088/0004-637X/780/1/57},
  \href {http://adsabs.harvard.edu/abs/2014ApJ...780...57B} {780, 57}

\bibitem[\protect\citeauthoryear{{Bournaud}, {Daddi}, {Wei{\ss}}, {Renaud},
  {Mastropietro}  \& {Teyssier}}{{Bournaud} et~al.}{2015}]{Bournaud:2015aa}
{Bournaud} F.,  {Daddi} E.,  {Wei{\ss}} A.,  {Renaud} F.,  {Mastropietro} C.,
  {Teyssier} R.,  2015, \mn@doi [\aap] {10.1051/0004-6361/201425078}, \href
  {http://adsabs.harvard.edu/abs/2015A%26A...575A..56B} {575, A56}

\bibitem[\protect\citeauthoryear{{Brown}, {Gnedin}  \& {Li}}{{Brown}
  et~al.}{2018}]{Brown:2018aa}
{Brown} G.,  {Gnedin} O.~Y.,   {Li} H.,  2018, \mn@doi [\apj]
  {10.3847/1538-4357/aad595}, \href
  {http://adsabs.harvard.edu/abs/2018ApJ...864...94B} {864, 94}

\bibitem[\protect\citeauthoryear{{Buck}, {Macci{\`o}}, {Obreja}, {Dutton},
  {Dom{\'{\i}}nguez-Tenreiro}  \& {Granato}}{{Buck} et~al.}{2017}]{Buck:2017aa}
{Buck} T.,  {Macci{\`o}} A.~V.,  {Obreja} A.,  {Dutton} A.~A.,
  {Dom{\'{\i}}nguez-Tenreiro} R.,   {Granato} G.~L.,  2017, \mn@doi [\mnras]
  {10.1093/mnras/stx685}, \href
  {http://adsabs.harvard.edu/abs/2017MNRAS.468.3628B} {468, 3628}

\bibitem[\protect\citeauthoryear{{Bullock}, {Kravtsov}  \&
  {Weinberg}}{{Bullock} et~al.}{2001}]{bullock_etal01}
{Bullock} J.~S.,  {Kravtsov} A.~V.,   {Weinberg} D.~H.,  2001, \apj, 548, 33

\bibitem[\protect\citeauthoryear{{Cava}, {Schaerer}, {Richard},
  {P{\'e}rez-Gonz{\'a}lez}, {Dessauges-Zavadsky}, {Mayer}  \&
  {Tamburello}}{{Cava} et~al.}{2018}]{Cava:2018aa}
{Cava} A.,  {Schaerer} D.,  {Richard} J.,  {P{\'e}rez-Gonz{\'a}lez} P.~G.,
  {Dessauges-Zavadsky} M.,  {Mayer} L.,   {Tamburello} V.,  2018, \mn@doi
  [Nature Astronomy] {10.1038/s41550-017-0295-x}, \href
  {http://adsabs.harvard.edu/abs/2018NatAs...2...76C} {2, 76}

\bibitem[\protect\citeauthoryear{{Ceverino}, {Dekel}  \& {Bournaud}}{{Ceverino}
  et~al.}{2010}]{Ceverino:2010aa}
{Ceverino} D.,  {Dekel} A.,   {Bournaud} F.,  2010, \mn@doi [\mnras]
  {10.1111/j.1365-2966.2010.16433.x}, \href
  {http://adsabs.harvard.edu/abs/2010MNRAS.404.2151C} {404, 2151}

\bibitem[\protect\citeauthoryear{{Ceverino}, {Dekel}, {Mandelker}, {Bournaud},
  {Burkert}, {Genzel}  \& {Primack}}{{Ceverino} et~al.}{2012}]{Ceverino:2012aa}
{Ceverino} D.,  {Dekel} A.,  {Mandelker} N.,  {Bournaud} F.,  {Burkert} A.,
  {Genzel} R.,   {Primack} J.,  2012, \mn@doi [\mnras]
  {10.1111/j.1365-2966.2011.20296.x}, \href
  {http://adsabs.harvard.edu/abs/2012MNRAS.420.3490C} {420, 3490}

\bibitem[\protect\citeauthoryear{{Conroy} \& {Gunn}}{{Conroy} \&
  {Gunn}}{2010}]{Conroy:2010aa}
{Conroy} C.,  {Gunn} J.~E.,  2010, \mn@doi [\apj]
  {10.1088/0004-637X/712/2/833}, \href
  {http://adsabs.harvard.edu/abs/2010ApJ...712..833C} {712, 833}

\bibitem[\protect\citeauthoryear{{Conroy}, {Gunn}  \& {White}}{{Conroy}
  et~al.}{2009}]{Conroy:2009aa}
{Conroy} C.,  {Gunn} J.~E.,   {White} M.,  2009, \mn@doi [\apj]
  {10.1088/0004-637X/699/1/486}, \href
  {http://adsabs.harvard.edu/abs/2009ApJ...699..486C} {699, 486}

\bibitem[\protect\citeauthoryear{{Dessauges-Zavadsky} \&
  {Adamo}}{{Dessauges-Zavadsky} \& {Adamo}}{2018}]{Dessauges-Zavadsky:2018aa}
{Dessauges-Zavadsky} M.,  {Adamo} A.,  2018, \mn@doi [\mnras]
  {10.1093/mnrasl/sly112}, \href
  {http://adsabs.harvard.edu/abs/2018MNRAS.479L.118D} {479, L118}

\bibitem[\protect\citeauthoryear{{Dessauges-Zavadsky}
  et~al.,}{{Dessauges-Zavadsky} et~al.}{2017a}]{Dessauges-Zavadsky:2017ab}
{Dessauges-Zavadsky} M.,  et~al., 2017a, \mn@doi [\aap]
  {10.1051/0004-6361/201628513}, \href
  {http://adsabs.harvard.edu/abs/2017A%26A...605A..81D} {605, A81}

\bibitem[\protect\citeauthoryear{{Dessauges-Zavadsky}, {Schaerer}, {Cava},
  {Mayer}  \& {Tamburello}}{{Dessauges-Zavadsky}
  et~al.}{2017b}]{Dessauges-Zavadsky:2017aa}
{Dessauges-Zavadsky} M.,  {Schaerer} D.,  {Cava} A.,  {Mayer} L.,
  {Tamburello} V.,  2017b, \mn@doi [\apjl] {10.3847/2041-8213/aa5d52}, \href
  {http://adsabs.harvard.edu/abs/2017ApJ...836L..22D} {836, L22}

\bibitem[\protect\citeauthoryear{{Elmegreen}, {Elmegreen}  \&
  {Sheets}}{{Elmegreen} et~al.}{2004}]{Elmegreen:2004ab}
{Elmegreen} D.~M.,  {Elmegreen} B.~G.,   {Sheets} C.~M.,  2004, \mn@doi [\apj]
  {10.1086/381357}, \href {http://adsabs.harvard.edu/abs/2004ApJ...603...74E}
  {603, 74}

\bibitem[\protect\citeauthoryear{{Elmegreen}, {Elmegreen}, {Rubin}  \&
  {Schaffer}}{{Elmegreen} et~al.}{2005}]{Elmegreen:2005aa}
{Elmegreen} D.~M.,  {Elmegreen} B.~G.,  {Rubin} D.~S.,   {Schaffer} M.~A.,
  2005, \mn@doi [\apj] {10.1086/432502}, \href
  {http://adsabs.harvard.edu/abs/2005ApJ...631...85E} {631, 85}

\bibitem[\protect\citeauthoryear{{Elmegreen}, {Elmegreen}, {Ravindranath}  \&
  {Coe}}{{Elmegreen} et~al.}{2007}]{Elmegreen:2007aa}
{Elmegreen} D.~M.,  {Elmegreen} B.~G.,  {Ravindranath} S.,   {Coe} D.~A.,
  2007, \mn@doi [\apj] {10.1086/511667}, \href
  {http://adsabs.harvard.edu/abs/2007ApJ...658..763E} {658, 763}

\bibitem[\protect\citeauthoryear{{Fisher} et~al.,}{{Fisher}
  et~al.}{2017a}]{Fisher:2017ab}
{Fisher} D.~B.,  et~al., 2017a, \mn@doi [\mnras] {10.1093/mnras/stw2281}, \href
  {http://adsabs.harvard.edu/abs/2017MNRAS.464..491F} {464, 491}

\bibitem[\protect\citeauthoryear{{Fisher} et~al.,}{{Fisher}
  et~al.}{2017b}]{Fisher:2017aa}
{Fisher} D.~B.,  et~al., 2017b, \mn@doi [\apjl] {10.3847/2041-8213/aa6478},
  \href {http://adsabs.harvard.edu/abs/2017ApJ...839L...5F} {839, L5}

\bibitem[\protect\citeauthoryear{{F{\"o}rster Schreiber} et~al.,}{{F{\"o}rster
  Schreiber} et~al.}{2006}]{Forster-Schreiber:2006aa}
{F{\"o}rster Schreiber} N.~M.,  et~al., 2006, \mn@doi [\apj] {10.1086/504403},
  \href {http://adsabs.harvard.edu/abs/2006ApJ...645.1062F} {645, 1062}

\bibitem[\protect\citeauthoryear{{F{\"o}rster Schreiber} et~al.,}{{F{\"o}rster
  Schreiber} et~al.}{2011}]{Forster-Schreiber:2011aa}
{F{\"o}rster Schreiber} N.~M.,  et~al., 2011, \mn@doi [\apj]
  {10.1088/0004-637X/739/1/45}, \href
  {http://adsabs.harvard.edu/abs/2011ApJ...739...45F} {739, 45}

\bibitem[\protect\citeauthoryear{{Genel} et~al.,}{{Genel}
  et~al.}{2012}]{Genel:2012aa}
{Genel} S.,  et~al., 2012, \mn@doi [\apj] {10.1088/0004-637X/745/1/11}, \href
  {http://adsabs.harvard.edu/abs/2012ApJ...745...11G} {745, 11}

\bibitem[\protect\citeauthoryear{{Genzel} et~al.,}{{Genzel}
  et~al.}{2008}]{Genzel:2008aa}
{Genzel} R.,  et~al., 2008, \mn@doi [\apj] {10.1086/591840}, \href
  {http://adsabs.harvard.edu/abs/2008ApJ...687...59G} {687, 59}

\bibitem[\protect\citeauthoryear{{Genzel} et~al.,}{{Genzel}
  et~al.}{2011}]{Genzel:2011aa}
{Genzel} R.,  et~al., 2011, \mn@doi [\apj] {10.1088/0004-637X/733/2/101}, \href
  {http://adsabs.harvard.edu/abs/2011ApJ...733..101G} {733, 101}

\bibitem[\protect\citeauthoryear{{Girard}, {Dessauges-Zavadsky}, {Schaerer},
  {Richard}, {Nakajima}  \& {Cava}}{{Girard} et~al.}{2018}]{Girard:2018aa}
{Girard} M.,  {Dessauges-Zavadsky} M.,  {Schaerer} D.,  {Richard} J.,
  {Nakajima} K.,   {Cava} A.,  2018, \mn@doi [\aap]
  {10.1051/0004-6361/201833533}, \href
  {http://adsabs.harvard.edu/abs/2018A%26A...619A..15G} {619, A15}

\bibitem[\protect\citeauthoryear{{Gnedin}}{{Gnedin}}{2014}]{Gnedin:2014aa}
{Gnedin} N.~Y.,  2014, \mn@doi [\apj] {10.1088/0004-637X/793/1/29}, \href
  {http://adsabs.harvard.edu/abs/2014ApJ...793...29G} {793, 29}

\bibitem[\protect\citeauthoryear{{Gnedin} \& {Abel}}{{Gnedin} \&
  {Abel}}{2001}]{Gnedin:2001aa}
{Gnedin} N.~Y.,  {Abel} T.,  2001, \mn@doi [NewA]
  {10.1016/S1384-1076(01)00068-9}, \href
  {http://adsabs.harvard.edu/abs/2001NewA....6..437G} {6, 437}

\bibitem[\protect\citeauthoryear{{Gnedin} \& {Kravtsov}}{{Gnedin} \&
  {Kravtsov}}{2011}]{Gnedin:2011aa}
{Gnedin} N.~Y.,  {Kravtsov} A.~V.,  2011, \mn@doi [\apj]
  {10.1088/0004-637X/728/2/88}, \href
  {http://adsabs.harvard.edu/abs/2011ApJ...728...88G} {728, 88}

\bibitem[\protect\citeauthoryear{{Guo}, {Giavalisco}, {Ferguson}, {Cassata}  \&
  {Koekemoer}}{{Guo} et~al.}{2012}]{Guo:2012aa}
{Guo} Y.,  {Giavalisco} M.,  {Ferguson} H.~C.,  {Cassata} P.,   {Koekemoer}
  A.~M.,  2012, \mn@doi [\apj] {10.1088/0004-637X/757/2/120}, \href
  {http://adsabs.harvard.edu/abs/2012ApJ...757..120G} {757, 120}

\bibitem[\protect\citeauthoryear{{Guo} et~al.,}{{Guo}
  et~al.}{2015}]{Guo:2015aa}
{Guo} Y.,  et~al., 2015, \mn@doi [\apj] {10.1088/0004-637X/800/1/39}, \href
  {http://adsabs.harvard.edu/abs/2015ApJ...800...39G} {800, 39}

\bibitem[\protect\citeauthoryear{{Guo} et~al.,}{{Guo}
  et~al.}{2018}]{Guo:2018aa}
{Guo} Y.,  et~al., 2018, \mn@doi [\apj] {10.3847/1538-4357/aaa018}, \href
  {http://adsabs.harvard.edu/abs/2018ApJ...853..108G} {853, 108}

\bibitem[\protect\citeauthoryear{{Haardt} \& {Madau}}{{Haardt} \&
  {Madau}}{2001}]{Haardt:2001aa}
{Haardt} F.,  {Madau} P.,  2001, Clusters of Galaxies and the High Redshift
  Universe Observed in X-rays, \href
  {http://adsabs.harvard.edu/abs/2001cghr.confE..64H} {p.~64}

\bibitem[\protect\citeauthoryear{{Hopkins}, {Kere{\v s}}, {Murray}, {Quataert}
  \& {Hernquist}}{{Hopkins} et~al.}{2012}]{Hopkins:2012aa}
{Hopkins} P.~F.,  {Kere{\v s}} D.,  {Murray} N.,  {Quataert} E.,   {Hernquist}
  L.,  2012, \mn@doi [\mnras] {10.1111/j.1365-2966.2012.21981.x}, \href
  {http://adsabs.harvard.edu/abs/2012MNRAS.427..968H} {427, 968}

\bibitem[\protect\citeauthoryear{{Inoue} \& {Saitoh}}{{Inoue} \&
  {Saitoh}}{2012}]{Inoue:2012aa}
{Inoue} S.,  {Saitoh} T.~R.,  2012, \mn@doi [\mnras]
  {10.1111/j.1365-2966.2011.20338.x}, \href
  {http://adsabs.harvard.edu/abs/2012MNRAS.422.1902I} {422, 1902}

\bibitem[\protect\citeauthoryear{{Inoue} \& {Yoshida}}{{Inoue} \&
  {Yoshida}}{2018}]{Inoue:2018aa}
{Inoue} S.,  {Yoshida} N.,  2018, \mn@doi [\mnras] {10.1093/mnras/stx2978},
  \href {http://adsabs.harvard.edu/abs/2018MNRAS.474.3466I} {474, 3466}

\bibitem[\protect\citeauthoryear{{Inoue}, {Dekel}, {Mandelker}, {Ceverino},
  {Bournaud}  \& {Primack}}{{Inoue} et~al.}{2016}]{Inoue:2016aa}
{Inoue} S.,  {Dekel} A.,  {Mandelker} N.,  {Ceverino} D.,  {Bournaud} F.,
  {Primack} J.,  2016, \mn@doi [\mnras] {10.1093/mnras/stv2793}, \href
  {http://adsabs.harvard.edu/abs/2016MNRAS.456.2052I} {456, 2052}

\bibitem[\protect\citeauthoryear{{Johnson} et~al.,}{{Johnson}
  et~al.}{2017}]{Johnson:2017ab}
{Johnson} T.~L.,  et~al., 2017, \mn@doi [\apjl] {10.3847/2041-8213/aa7516},
  \href {http://adsabs.harvard.edu/abs/2017ApJ...843L..21J} {843, L21}

\bibitem[\protect\citeauthoryear{{Jones}, {Swinbank}, {Ellis}, {Richard}  \&
  {Stark}}{{Jones} et~al.}{2010}]{Jones:2010aa}
{Jones} T.~A.,  {Swinbank} A.~M.,  {Ellis} R.~S.,  {Richard} J.,   {Stark}
  D.~P.,  2010, \mn@doi [\mnras] {10.1111/j.1365-2966.2010.16378.x}, \href
  {http://adsabs.harvard.edu/abs/2010MNRAS.404.1247J} {404, 1247}

\bibitem[\protect\citeauthoryear{{Kennicutt}}{{Kennicutt}}{1998}]{Kennicutt:1998aa}
{Kennicutt} Robert~C. J.,  1998, \mn@doi [\araa]
  {10.1146/annurev.astro.36.1.189}, \href
  {https://ui.adsabs.harvard.edu/abs/1998ARA&A..36..189K} {36, 189}

\bibitem[\protect\citeauthoryear{{Kennicutt} \& {Evans}}{{Kennicutt} \&
  {Evans}}{2012}]{Kennicutt:2012aa}
{Kennicutt} R.~C.,  {Evans} N.~J.,  2012, \mn@doi [\araa]
  {10.1146/annurev-astro-081811-125610}, \href
  {https://ui.adsabs.harvard.edu/abs/2012ARA&A..50..531K} {50, 531}

\bibitem[\protect\citeauthoryear{{Kravtsov}}{{Kravtsov}}{1999}]{Kravtsov:1999aa}
{Kravtsov} A.~V.,  1999, PhD thesis, NEW MEXICO STATE UNIVERSITY

\bibitem[\protect\citeauthoryear{{Kravtsov}}{{Kravtsov}}{2003}]{Kravtsov:2003aa}
{Kravtsov} A.~V.,  2003, \mn@doi [\apjl] {10.1086/376674}, \href
  {http://adsabs.harvard.edu/abs/2003ApJ...590L...1K} {590, L1}

\bibitem[\protect\citeauthoryear{{Kravtsov}, {Klypin}  \&
  {Khokhlov}}{{Kravtsov} et~al.}{1997}]{Kravtsov:1997aa}
{Kravtsov} A.~V.,  {Klypin} A.~A.,   {Khokhlov} A.~M.,  1997, \mn@doi [\apjs]
  {10.1086/313015}, \href {http://adsabs.harvard.edu/abs/1997ApJS..111...73K}
  {111, 73}

\bibitem[\protect\citeauthoryear{{Kroupa}}{{Kroupa}}{2001}]{Kroupa:2001aa}
{Kroupa} P.,  2001, \mn@doi [\mnras] {10.1046/j.1365-8711.2001.04022.x}, \href
  {https://ui.adsabs.harvard.edu/abs/2001MNRAS.322..231K} {322, 231}

\bibitem[\protect\citeauthoryear{{Li}, {Gnedin}, {Gnedin}, {Meng}, {Semenov}
  \& {Kravtsov}}{{Li} et~al.}{2017}]{Li:2017ab}
{Li} H.,  {Gnedin} O.~Y.,  {Gnedin} N.~Y.,  {Meng} X.,  {Semenov} V.~A.,
  {Kravtsov} A.~V.,  2017, \mn@doi [\apj] {10.3847/1538-4357/834/1/69}, \href
  {http://adsabs.harvard.edu/abs/2017ApJ...834...69L} {834, 69}

\bibitem[\protect\citeauthoryear{{Li}, {Gnedin}  \& {Gnedin}}{{Li}
  et~al.}{2018}]{Li:2018aa}
{Li} H.,  {Gnedin} O.~Y.,   {Gnedin} N.~Y.,  2018, \mn@doi [\apj]
  {10.3847/1538-4357/aac9b8}, \href
  {http://adsabs.harvard.edu/abs/2018ApJ...861..107L} {861, 107}

\bibitem[\protect\citeauthoryear{{Livermore} et~al.,}{{Livermore}
  et~al.}{2012}]{Livermore:2012aa}
{Livermore} R.~C.,  et~al., 2012, \mn@doi [\mnras]
  {10.1111/j.1365-2966.2012.21900.x}, \href
  {http://adsabs.harvard.edu/abs/2012MNRAS.427..688L} {427, 688}

\bibitem[\protect\citeauthoryear{{Livermore} et~al.,}{{Livermore}
  et~al.}{2015}]{Livermore:2015aa}
{Livermore} R.~C.,  et~al., 2015, \mn@doi [\mnras] {10.1093/mnras/stv686},
  \href {http://adsabs.harvard.edu/abs/2015MNRAS.450.1812L} {450, 1812}

\bibitem[\protect\citeauthoryear{{Madau} \& {Dickinson}}{{Madau} \&
  {Dickinson}}{2014}]{Madau:2014aa}
{Madau} P.,  {Dickinson} M.,  2014, \mn@doi [\araa]
  {10.1146/annurev-astro-081811-125615}, \href
  {http://adsabs.harvard.edu/abs/2014ARA%26A..52..415M} {52, 415}

\bibitem[\protect\citeauthoryear{{Mandelker}, {Dekel}, {Ceverino}, {Tweed},
  {Moody}  \& {Primack}}{{Mandelker} et~al.}{2014}]{Mandelker:2014aa}
{Mandelker} N.,  {Dekel} A.,  {Ceverino} D.,  {Tweed} D.,  {Moody} C.~E.,
  {Primack} J.,  2014, \mn@doi [\mnras] {10.1093/mnras/stu1340}, \href
  {http://adsabs.harvard.edu/abs/2014MNRAS.443.3675M} {443, 3675}

\bibitem[\protect\citeauthoryear{{Mandelker}, {Dekel}, {Ceverino}, {DeGraf},
  {Guo}  \& {Primack}}{{Mandelker} et~al.}{2017}]{Mandelker:2017aa}
{Mandelker} N.,  {Dekel} A.,  {Ceverino} D.,  {DeGraf} C.,  {Guo} Y.,
  {Primack} J.,  2017, \mn@doi [\mnras] {10.1093/mnras/stw2358}, \href
  {http://adsabs.harvard.edu/abs/2017MNRAS.464..635M} {464, 635}

\bibitem[\protect\citeauthoryear{{Martizzi}, {Faucher-Gigu{\`e}re}  \&
  {Quataert}}{{Martizzi} et~al.}{2015}]{Martizzi:2015aa}
{Martizzi} D.,  {Faucher-Gigu{\`e}re} C.-A.,   {Quataert} E.,  2015, \mn@doi
  [\mnras] {10.1093/mnras/stv562}, \href
  {http://adsabs.harvard.edu/abs/2015MNRAS.450..504M} {450, 504}

\bibitem[\protect\citeauthoryear{{Mayer}, {Tamburello}, {Lupi}, {Keller},
  {Wadsley}  \& {Madau}}{{Mayer} et~al.}{2016}]{Mayer:2016aa}
{Mayer} L.,  {Tamburello} V.,  {Lupi} A.,  {Keller} B.,  {Wadsley} J.,
  {Madau} P.,  2016, \mn@doi [\apjl] {10.3847/2041-8205/830/1/L13}, \href
  {http://adsabs.harvard.edu/abs/2016ApJ...830L..13M} {830, L13}

\bibitem[\protect\citeauthoryear{{Meng}, {Gnedin}  \& {Li}}{{Meng}
  et~al.}{2019}]{Meng:2019aa}
{Meng} X.,  {Gnedin} O.~Y.,   {Li} H.,  2019, \mn@doi [\mnras]
  {10.1093/mnras/stz925}, \href
  {https://ui.adsabs.harvard.edu/abs/2019MNRAS.486.1574M} {486, 1574}

\bibitem[\protect\citeauthoryear{{Mieda}, {Wright}, {Larkin}, {Armus},
  {Juneau}, {Salim}  \& {Murray}}{{Mieda} et~al.}{2016}]{Mieda:2016aa}
{Mieda} E.,  {Wright} S.~A.,  {Larkin} J.~E.,  {Armus} L.,  {Juneau} S.,
  {Salim} S.,   {Murray} N.,  2016, \mn@doi [\apj]
  {10.3847/0004-637X/831/1/78}, \href
  {http://adsabs.harvard.edu/abs/2016ApJ...831...78M} {831, 78}

\bibitem[\protect\citeauthoryear{{Moody}, {Guo}, {Mandelker}, {Ceverino},
  {Mozena}, {Koo}, {Dekel}  \& {Primack}}{{Moody} et~al.}{2014}]{Moody:2014aa}
{Moody} C.~E.,  {Guo} Y.,  {Mandelker} N.,  {Ceverino} D.,  {Mozena} M.,  {Koo}
  D.~C.,  {Dekel} A.,   {Primack} J.,  2014, \mn@doi [\mnras]
  {10.1093/mnras/stu1534}, \href
  {http://adsabs.harvard.edu/abs/2014MNRAS.444.1389M} {444, 1389}

\bibitem[\protect\citeauthoryear{{Murata} et~al.,}{{Murata}
  et~al.}{2014}]{Murata:2014aa}
{Murata} K.~L.,  et~al., 2014, \mn@doi [\apj] {10.1088/0004-637X/786/1/15},
  \href {http://adsabs.harvard.edu/abs/2014ApJ...786...15M} {786, 15}

\bibitem[\protect\citeauthoryear{{Murray}, {Quataert}  \& {Thompson}}{{Murray}
  et~al.}{2010}]{Murray:2010aa}
{Murray} N.,  {Quataert} E.,   {Thompson} T.~A.,  2010, \mn@doi [\apj]
  {10.1088/0004-637X/709/1/191}, \href
  {http://adsabs.harvard.edu/abs/2010ApJ...709..191M} {709, 191}

\bibitem[\protect\citeauthoryear{{Noguchi}}{{Noguchi}}{1999}]{Noguchi:1999aa}
{Noguchi} M.,  1999, \mn@doi [\apj] {10.1086/306932}, \href
  {http://adsabs.harvard.edu/abs/1999ApJ...514...77N} {514, 77}

\bibitem[\protect\citeauthoryear{{Oklop{\v c}i{\'c}}, {Hopkins}, {Feldmann},
  {Kere{\v s}}, {Faucher-Gigu{\`e}re}  \& {Murray}}{{Oklop{\v c}i{\'c}}
  et~al.}{2017}]{Oklopcic:2017aa}
{Oklop{\v c}i{\'c}} A.,  {Hopkins} P.~F.,  {Feldmann} R.,  {Kere{\v s}} D.,
  {Faucher-Gigu{\`e}re} C.-A.,   {Murray} N.,  2017, \mn@doi [\mnras]
  {10.1093/mnras/stw2754}, \href
  {http://adsabs.harvard.edu/abs/2017MNRAS.465..952O} {465, 952}

\bibitem[\protect\citeauthoryear{{Olmstead}, {Rigby}, {Swinbank}  \&
  {Veilleux}}{{Olmstead} et~al.}{2014}]{Olmstead:2014aa}
{Olmstead} A.,  {Rigby} J.~R.,  {Swinbank} M.,   {Veilleux} S.,  2014, \mn@doi
  [\aj] {10.1088/0004-6256/148/4/65}, \href
  {http://adsabs.harvard.edu/abs/2014AJ....148...65O} {148, 65}

\bibitem[\protect\citeauthoryear{{Overzier}, {Heckman}, {Schiminovich},
  {Basu-Zych}, {Gon{\c c}alves}, {Martin}  \& {Rich}}{{Overzier}
  et~al.}{2010}]{Overzier:2010aa}
{Overzier} R.~A.,  {Heckman} T.~M.,  {Schiminovich} D.,  {Basu-Zych} A.,
  {Gon{\c c}alves} T.,  {Martin} D.~C.,   {Rich} R.~M.,  2010, \mn@doi [\apj]
  {10.1088/0004-637X/710/2/979}, \href
  {http://adsabs.harvard.edu/abs/2010ApJ...710..979O} {710, 979}

\bibitem[\protect\citeauthoryear{{Perez}, {Valenzuela}, {Tissera}  \&
  {Michel-Dansac}}{{Perez} et~al.}{2013}]{Perez:2013aa}
{Perez} J.,  {Valenzuela} O.,  {Tissera} P.~B.,   {Michel-Dansac} L.,  2013,
  \mn@doi [\mnras] {10.1093/mnras/stt1563}, \href
  {http://adsabs.harvard.edu/abs/2013MNRAS.436..259P} {436, 259}

\bibitem[\protect\citeauthoryear{{Puech}}{{Puech}}{2010}]{Puech:2010aa}
{Puech} M.,  2010, \mn@doi [\mnras] {10.1111/j.1365-2966.2010.16689.x}, \href
  {http://adsabs.harvard.edu/abs/2010MNRAS.406..535P} {406, 535}

\bibitem[\protect\citeauthoryear{{Puech}, {Hammer}, {Flores}, {Neichel}  \&
  {Yang}}{{Puech} et~al.}{2009}]{Puech:2009aa}
{Puech} M.,  {Hammer} F.,  {Flores} H.,  {Neichel} B.,   {Yang} Y.,  2009,
  \mn@doi [\aap] {10.1051/0004-6361:200810521}, \href
  {http://adsabs.harvard.edu/abs/2009A%26A...493..899P} {493, 899}

\bibitem[\protect\citeauthoryear{{Ribeiro} et~al.,}{{Ribeiro}
  et~al.}{2017}]{Ribeiro:2017aa}
{Ribeiro} B.,  et~al., 2017, \mn@doi [\aap] {10.1051/0004-6361/201630057},
  \href {http://adsabs.harvard.edu/abs/2017A%26A...608A..16R} {608, A16}

\bibitem[\protect\citeauthoryear{{Rudd}, {Zentner}  \& {Kravtsov}}{{Rudd}
  et~al.}{2008}]{Rudd:2008aa}
{Rudd} D.~H.,  {Zentner} A.~R.,   {Kravtsov} A.~V.,  2008, \mn@doi [\apj]
  {10.1086/523836}, \href {http://adsabs.harvard.edu/abs/2008ApJ...672...19R}
  {672, 19}

\bibitem[\protect\citeauthoryear{{Rujopakarn} et~al.,}{{Rujopakarn}
  et~al.}{2019}]{Rujopakarn:2019aa}
{Rujopakarn} W.,  et~al., 2019, \mn@doi [\apj] {10.3847/1538-4357/ab3791},
  \href {https://ui.adsabs.harvard.edu/abs/2019ApJ...882..107R} {882, 107}

\bibitem[\protect\citeauthoryear{{Safronov}}{{Safronov}}{1960}]{safronov60}
{Safronov} V.~S.,  1960, Annales d'Astrophysique, \href
  {http://adsabs.harvard.edu/abs/1960AnAp...23..979S} {23, 979}

\bibitem[\protect\citeauthoryear{{Schmidt} et~al.,}{{Schmidt}
  et~al.}{2014}]{Schmidt:2014aa}
{Schmidt} W.,  et~al., 2014, \mn@doi [\mnras] {10.1093/mnras/stu501}, \href
  {http://adsabs.harvard.edu/abs/2014MNRAS.440.3051S} {440, 3051}

\bibitem[\protect\citeauthoryear{{Semenov}, {Kravtsov}  \& {Gnedin}}{{Semenov}
  et~al.}{2016}]{Semenov:2016aa}
{Semenov} V.~A.,  {Kravtsov} A.~V.,   {Gnedin} N.~Y.,  2016, \mn@doi [\apj]
  {10.3847/0004-637X/826/2/200}, \href
  {http://adsabs.harvard.edu/abs/2016ApJ...826..200S} {826, 200}

\bibitem[\protect\citeauthoryear{{Shapiro} et~al.,}{{Shapiro}
  et~al.}{2008}]{Shapiro:2008aa}
{Shapiro} K.~L.,  et~al., 2008, \mn@doi [\apj] {10.1086/587133}, \href
  {http://adsabs.harvard.edu/abs/2008ApJ...682..231S} {682, 231}

\bibitem[\protect\citeauthoryear{{Shibuya}, {Ouchi}, {Kubo}  \&
  {Harikane}}{{Shibuya} et~al.}{2016}]{Shibuya:2016aa}
{Shibuya} T.,  {Ouchi} M.,  {Kubo} M.,   {Harikane} Y.,  2016, \mn@doi [\apj]
  {10.3847/0004-637X/821/2/72}, \href
  {http://adsabs.harvard.edu/abs/2016ApJ...821...72S} {821, 72}

\bibitem[\protect\citeauthoryear{{Soto} et~al.,}{{Soto}
  et~al.}{2017}]{Soto:2017aa}
{Soto} E.,  et~al., 2017, \mn@doi [\apj] {10.3847/1538-4357/aa5da3}, \href
  {http://adsabs.harvard.edu/abs/2017ApJ...837....6S} {837, 6}

\bibitem[\protect\citeauthoryear{{Spitzer}}{{Spitzer}}{1969}]{Spitzer:1969aa}
{Spitzer} Jr. L.,  1969, \mn@doi [\apjl] {10.1086/180451}, \href
  {https://ui.adsabs.harvard.edu/abs/1969ApJ...158L.139S} {158, L139}

\bibitem[\protect\citeauthoryear{{Swinbank} et~al.,}{{Swinbank}
  et~al.}{2010a}]{Swinbank:2010aa}
{Swinbank} A.~M.,  et~al., 2010a, \mn@doi [\mnras]
  {10.1111/j.1365-2966.2010.16485.x}, \href
  {http://adsabs.harvard.edu/abs/2010MNRAS.405..234S} {405, 234}

\bibitem[\protect\citeauthoryear{{Swinbank} et~al.,}{{Swinbank}
  et~al.}{2010b}]{Swinbank:2010ab}
{Swinbank} A.~M.,  et~al., 2010b, \mn@doi [\nat] {10.1038/nature08880}, \href
  {http://adsabs.harvard.edu/abs/2010Natur.464..733S} {464, 733}

\bibitem[\protect\citeauthoryear{{Tadaki}, {Kodama}, {Tanaka}, {Hayashi},
  {Koyama}  \& {Shimakawa}}{{Tadaki} et~al.}{2014}]{Tadaki:2014aa}
{Tadaki} K.-i.,  {Kodama} T.,  {Tanaka} I.,  {Hayashi} M.,  {Koyama} Y.,
  {Shimakawa} R.,  2014, \mn@doi [\apj] {10.1088/0004-637X/780/1/77}, \href
  {http://adsabs.harvard.edu/abs/2014ApJ...780...77T} {780, 77}

\bibitem[\protect\citeauthoryear{{Tamburello}, {Mayer}, {Shen}  \&
  {Wadsley}}{{Tamburello} et~al.}{2015}]{Tamburello:2015aa}
{Tamburello} V.,  {Mayer} L.,  {Shen} S.,   {Wadsley} J.,  2015, \mn@doi
  [\mnras] {10.1093/mnras/stv1695}, \href
  {http://adsabs.harvard.edu/abs/2015MNRAS.453.2490T} {453, 2490}

\bibitem[\protect\citeauthoryear{{Tamburello}, {Rahmati}, {Mayer}, {Cava},
  {Dessauges-Zavadsky}  \& {Schaerer}}{{Tamburello}
  et~al.}{2017}]{Tamburello:2017aa}
{Tamburello} V.,  {Rahmati} A.,  {Mayer} L.,  {Cava} A.,  {Dessauges-Zavadsky}
  M.,   {Schaerer} D.,  2017, \mn@doi [\mnras] {10.1093/mnras/stx784}, \href
  {http://adsabs.harvard.edu/abs/2017MNRAS.468.4792T} {468, 4792}

\bibitem[\protect\citeauthoryear{{Toomre}}{{Toomre}}{1964}]{toomre64}
{Toomre} A.,  1964, \mn@doi [\apj] {10.1086/147861}, \href
  {http://adsabs.harvard.edu/abs/1964ApJ...139.1217T} {139, 1217}

\bibitem[\protect\citeauthoryear{{Wisnioski}, {Glazebrook}, {Blake}, {Poole},
  {Green}, {Wyder}  \& {Martin}}{{Wisnioski} et~al.}{2012}]{Wisnioski:2012aa}
{Wisnioski} E.,  {Glazebrook} K.,  {Blake} C.,  {Poole} G.~B.,  {Green} A.~W.,
  {Wyder} T.,   {Martin} C.,  2012, \mn@doi [\mnras]
  {10.1111/j.1365-2966.2012.20850.x}, \href
  {https://ui.adsabs.harvard.edu/abs/2012MNRAS.422.3339W} {422, 3339}

\bibitem[\protect\citeauthoryear{{Wisnioski} et~al.,}{{Wisnioski}
  et~al.}{2015}]{Wisnioski:2015aa}
{Wisnioski} E.,  et~al., 2015, \mn@doi [\apj] {10.1088/0004-637X/799/2/209},
  \href {http://adsabs.harvard.edu/abs/2015ApJ...799..209W} {799, 209}

\bibitem[\protect\citeauthoryear{{Wuyts} et~al.,}{{Wuyts}
  et~al.}{2012}]{Wuyts:2012aa}
{Wuyts} S.,  et~al., 2012, \mn@doi [\apj] {10.1088/0004-637X/753/2/114}, \href
  {http://adsabs.harvard.edu/abs/2012ApJ...753..114W} {753, 114}

\bibitem[\protect\citeauthoryear{{Wuyts}, {Rigby}, {Gladders}  \&
  {Sharon}}{{Wuyts} et~al.}{2014}]{Wuyts:2014aa}
{Wuyts} E.,  {Rigby} J.~R.,  {Gladders} M.~D.,   {Sharon} K.,  2014, \mn@doi
  [\apj] {10.1088/0004-637X/781/2/61}, \href
  {http://adsabs.harvard.edu/abs/2014ApJ...781...61W} {781, 61}

\bibitem[\protect\citeauthoryear{{Zanella} et~al.,}{{Zanella}
  et~al.}{2015}]{Zanella:2015aa}
{Zanella} A.,  et~al., 2015, \mn@doi [\nat] {10.1038/nature14409}, \href
  {http://adsabs.harvard.edu/abs/2015Natur.521...54Z} {521, 54}

\bibitem[\protect\citeauthoryear{{Zanella} et~al.,}{{Zanella}
  et~al.}{2019}]{Zanella:2019aa}
{Zanella} A.,  et~al., 2019, \mn@doi [\mnras] {10.1093/mnras/stz2099}, \href
  {https://ui.adsabs.harvard.edu/abs/2019MNRAS.489.2792Z} {489, 2792}

\makeatother
\end{thebibliography}

\bsp	
\label{lastpage}
\end{document}